\documentclass[amsmath, amsfonts, superscriptaddress, showpacs, twocolumn, prb]{revtex4}
\usepackage{graphicx}
\usepackage{epsfig}
\usepackage{bm}
\usepackage{dcolumn}
\usepackage{amsmath}
\usepackage{amssymb}
\usepackage{color}
\usepackage[varg]{txfonts}

\begin{document}

\title{Onset of surface superconductivity beyond the Saint-James--de Gennes limit}

\author{Hong-Yi Xie}
\affiliation{Department of Physics, University of Wisconsin-Madison, Madison, Wisconsin 53706, USA}

\author{Vladimir G. Kogan}
\affiliation{Ames Laboratory, DOE and Department of Physics, Iowa State University, Ames, Iowa 50011, USA}

\author{Maxim Khodas}
\affiliation{Racah Institute of Physics, Hebrew University of Jerusalem, Jerusalem 91904, Israel}

\author{Alex Levchenko}
\affiliation{Department of Physics, University of Wisconsin-Madison, Madison, Wisconsin 53706, USA}

\begin{abstract}
We revisit the problem of the surface superconductivity nucleation focusing on the detailed study of the critical field $H_{c3}$ as a function of temperature and disorder. Using the semiclassical Eilenberger formalism we find that away from the Ginzburg-Landau region the ratio between the nucleation critical field $H_{c3}$ and the upper critical field $H_{c2}$ deviates strongly from the Saint-James--de Gennes limit. In particular, the $H_{c3}/H_{c2}$ is found to be a nonmonotonic function of temperature, which reaches the maximum for a set of parameters corresponding to a crossover region from ballistic to diffusive scattering, when the mean-free path in a bulk of a superconductor is of the same order as zero-temperature superconducting coherence length. We also analyze the robustness of the nucleated phases with respect to diffusive scattering off the sample boundary by solving exactly corresponding eigenvalue problem of an integral equation for the critical field. The implications of these results for the transport in superconductors of various geometries near $H_{c3}$ are briefly discussed. In particular, we present results for the mechanism of magnetoconductivity oscillations due to  surface superconductivity effects.     
\end{abstract}

\date{September 25, 2017}

\pacs{74.25.Op, 74.62.En, 74.78.Na}

\maketitle

\section{Introduction}

The magnetic phase diagram of a type-II superconductor is marked by three distinct critical lines. The Meissner phase occurs at the lowest fields and extends up to the first critical field $H_{c1}$ that marks vortex entry into a superconductor to be energetically favorable. Next, the Abrikosov-vortex-lattice phase emerges and persists up to the second critical field $H_{c2}$. As vortex lattice becomes so dense that vortex cores overlap, superconductivity becomes extinguished from the bulk, yet it can nucleate at the surfaces and Saint-James--de Gennes (SJdG) phase of surface dominated superconductivity survives to even higher fields marked by $H_{c3}$. An exact solution of the linearized Ginzburg-Landau (GL) equations in the presence of the surface lead Saint-James--de Gennes to a famous result $H_{c3}=1.695H_{c2}$ [\onlinecite{SJdG}]. Even though the validity of this formula is limited to vicinity of the critical temperature $T_c$, still SJdG theory provided a rationale for explaining a great amount of experimental data on the persistence of superconductivity at high fields which had initially been ignored and attributed to inhomogeneities of samples under study [\onlinecite{Saint-James}]. 

The early evidence for surface superconductivity was found in niobium, lead, tin, and indium via various experiments including dc transport, susceptibility and magnetization measurements, torque magnetometry, surface impedance, and tunneling spectroscopy probes [\onlinecite{Tomasch,Hempstead,Strongin-PRL64,Strongin-PRL65,Schweitzer,Rothwarf,McEvoy,Cruz}]. Subsequent more detailed experiments [\onlinecite{Rollins,Hopkins-PRB74,Kirschenbaum-PRB95}] carried out on Pb-In and Nb-Ta alloys or oxidized Nb addressed dependence of $H_{c3}$ on temperature $T$ and sample purity in a broad range of parameters away from the validity region of the GL theory. It was revealed that the $H_{c3}/H_{c2}$ ratio is not a universal number but rather a complicated function that exhibits pronounced dependence on both $T$ and the mean-free time $\tau$ mediated by disorder scattering. Niobium became essentially a paradigmatic material to study various phenomena associated with surface superconductivity and multiple experiments continue to follow [\onlinecite{Park-PRL03,Stamopoulos-PRB04,Scola-PRB05,Das-PRB08,Werner-SST,Krasnov-PRB16,Kozhevnikov-arXiv17}]. Naturally over time experimental efforts on surface superconductivity branched into different directions. One line of interest shifted towards investigation of the effects of sample geometry and topology on nucleation of superconductivity and corresponding values of the critical field [\onlinecite{Moshchalkov,Bruyndoncx,Chandrasekhar}]. Other interests moved towards studying different superconducting compounds including: yttrium hexaboride YB$_6$ [\onlinecite{Tsindlekht-JPCM10}], multiband superconducting materials such as NbSe$_2$ [\onlinecite{DAnna-PRB96}] and MgB$_2$ [\onlinecite{Rydh-PRB03,Tsindlekht-PRB06}], unconventional superconductors such as heavy-fermion systems UPt$_3$ [\onlinecite{Keller-PRB96}] and iron-pnictides K$_x$Fe$_{2-x}$Se$_2$ [\onlinecite{Tsindlekht-PRB11}], and most recently layered dichalcogenide Cu$_x$TiSe$_2$ [\onlinecite{Levy-Bertrand}]. 

The original study of SJdG triggered not only a flurry of experimental investigations, but obviously also attracted substantial theoretical attention that lead to multiple developments and generalizations. Abrikosov was perhaps the first who looked at the problem of surface superconductivity beyond the framework of Ginzburg-Landau phenomenology [\onlinecite{Abrikosov}]. He adopted Gor'kov's method of computing $H_{c2}$ at zero temperature [\onlinecite{Gorkov}] and generalized it to the case of $H_{c3}$. Technically, this amounts to solving an eigenvalue problem for the linearized gap equation, which is an integral equation. Abrikosov commented in the paper that it is unusually difficult to evaluate the critical field of nucleation in the general case and, based on the experimental evidence available at that time, conjectured that SJdG relation between $H_{c3}$ and $H_{c2}$ will be satisfied over the entire range of temperatures below $T_c$. Hu and Korenman [\onlinecite{Hu-Korenman}] revisited this problem focusing on the simplest case of a clean superconductor with specular reflection from the boundary. They applied variational ansatz to tackle the integral eigenvalue problem for $H_{c3}$ at $T=0$ and obtained lower- and upper-bound estimates for the ratio between critical fields to be in the range $1.925 \leq H_{c3}/H_{c2}\leq 5.22$. The impact of the sample geometry was analyzed by van Gelder [\onlinecite{Gelder}] who predicted that for superconducting wedges subtending an angle $\gamma$ the surface critical field $H_{c3}(\gamma)$ can be greatly enhanced, thus easily overcoming the SJdG threshold corresponding to $\gamma=\pi$. These early theories were extended to investigate signatures of surface superconductivity fluctuations on transport coefficients [\onlinecite{Schmidt,Zyuzin}] and thermodynamic properties [\onlinecite{Aleiner}] at the onset of nucleation, and further to analyze critical field in the superconducting films focusing specifically on the boundary and finite thickness effects [\onlinecite{Scotto,Hara}]. The effect of anisotropy on $H_{c3}$ was considered within GL theory [\onlinecite{Kogan-Hc3}]. The critical field was also analyzed for the case of unconventional pairing [\onlinecite{Samokhin,Agterberg}], and for the case of multiband superconductor in the context of MgB$_2$ assuming disordered dominated regimes that were treated with the help of semiclassical Usadel equations [\onlinecite{Gorokhov}]. 

In light of all the existing studies, it is perhaps surprising to realize that we still do not have comprehensive solution of the surface superconductivity problem in terms of the corresponding critical field dependence on 
the broad range of essential parameters such as temperature $T$ and sample purity characterized by a mean-free time $\tau$. This situation should be contrasted to the case of $H_{c2}(T,\tau)$ studies for which Helfand and Werthamer provided a complete solution [\onlinecite{HW}]. While $H_{c2}$ is certainly a fundamentally important metric characterizing superconducting state in the bulk, the field of nucleation $H_{c3}$ serves as a benchmark for the mesoscale surface superconductivity. It is our primary motivation to generalize Helfand-Werthamer (HW) treatment to the case of $H_{c3}(T,\tau)$. This task is accomplished in Sec. \ref{Sec:Eilenberger-Hc3}. Additionally, in Sec. \ref{Sec:Boundary-SSC}  we also address the fate of surface superconductivity for rough surfaces that induce diffusive scattering off the boundary. In contrast to a naive expectation that poor quality of the surface should weaken $H_{c3}$ and strongly suppress surface superconductivity, we find that the magnitude of $H_{c3}$ in this case remains above the threshold of the SJdG limit. In Sec. \ref{Sec:Hc3-Disc} we study $H_{c3}$ for systems of cylindrical geometry that are highly relevant to numerous experiments conducted on superconducting nano-islands, rings, and quantum wires. In particular, we address scanning tunneling spectroscopy measurements of vortex trapping. Finally, in Sec. \ref{Sec:Transport} we comment on manifestations of surface conductivity in transport and discuss several devices that could reveal novel interesting aspects of this general phenomenon. Specifically, we discuss a mechanism for Aharonov-Bohm oscillations in magnetoconductivity as promoted by surface superconductivity effects. We summarize our findings in Sec. \ref{Sec:SumDis} and outline possible interesting questions for further exploration of surface superconductivity. 

\section{Eilenberger formalism for $H_{c3}$}\label{Sec:Eilenberger-Hc3}

The main idea of the semiclassical formalism, that was developed by Eilenberger in applications to the problems of superconductivity [\onlinecite{Eilenberger}], is to take advantage of the fact that length scales (e.g., coherence length or magnetic penetration depth) at which characteristic properties of a superconductor change, are large compared to electron Fermi wave length $\lambda_F$. This allows one to integrate exact Green's functions $G$ and $F$ in the theory of superconductivity over the energy variable. This step simplifies Gor'kov's equations to technically easier equations for the reduced semiclassical normal-$g(\bm{r},\omega,\bm{v})$ and anomalous-$f(\bm{r},\omega,\bm{v})$ Green's functions. In particular, the equation for the latter reads as (hereafter $\hbar=k_B=c=1$)
\begin{equation}\label{EE}
(2\omega+\bm{v}\bm{\Pi})f=2\Delta g+\frac{1}{\tau}\left(g\langle f\rangle-f \langle g\rangle\right).
\end{equation}
Here, $\omega=\pi T(2n+1)$ is the fermionic Matsubara frequency with $n=0,1,2\,\ldots$, the momentum operator contains vector potential $\bm{\Pi}=\bm{\nabla}+2\pi i\bm{A}/\phi_0$, $\phi_0=\pi/|e|$ is the flux quantum, and $\tau$ is the elastic scattering time due to non-magnetic impurities, whereas angular brackets $\langle\ldots\rangle$ denote the average over the Fermi surface (or over the all directions of velocity $\bm{v}$). The corresponding equation for the $g$-function is similar. As is clear, Eilenberger equation \eqref{EE} is intrinsically nonlinear, furthermore, Green's functions are subject to the normalization constraint $g^2+ff^\dag=1$, where $f^\dag(\bm{r},\omega,\bm{v})=f^*(\bm{r},\omega,-\bm{v})$. The superconducting pair-potential $\Delta(\bm{r})$ should be found from the self-consistency equation, that reads as      
\begin{equation}\label{Delta}
\Delta\ln\frac{T_c}{T}=2\pi T\sum_{\omega>0}\left[\frac{\Delta}{\omega}-\langle f\rangle\right]\, , 
\end{equation}
where $T_c$ is the transition temperature at zero field.
There are a number of physically interesting situations when the general Eilenberger equation can be further simplified. In particular, at the onset of the second-order superconductor-normal phase transition the pair potential is small so that Eilenberger equation can be linearized: $\tau\bm{v}\bm{\Pi}f=2\Delta\tau+\langle f\rangle-(1+2\omega\tau)f$. It has been shown [\onlinecite{Kogan-PRB85,Kogan-PRB87}] that solving such linearized equation with the self-consistency condition \eqref{Delta} is equivalent to solving an eigenvalue problem for the linear differential equation
\begin{equation}\label{GL}
\Pi^2\Delta(\bm{r})=k^2\Delta(\bm{r}),
\end{equation}  
where eigenvalue $k^2$, that defines superconducting coherence length $\xi^2=-1/k^2$ at any temperature $T$, magnetic field $H$, and scattering rate $\tau$, should be found from 
\begin{equation}\label{SC-Eq}
\ln\frac{T_c}{T}=2\pi T\sum_{\omega>0}\left(\frac{1}{\omega}-\frac{2\tau S}{\beta-S}\right),\quad \beta=1+2\omega\tau.
\end{equation}
Here the new function $S$ admits the following integral representation:
\begin{equation}\label{S}
S=\frac{\sqrt{\pi}}{\alpha}\int^{1}_{0}dy\frac{(1+y^2)^\sigma}{(1-y^2)^{\sigma+1}}\left[\Phi\left(\frac{y}{\alpha}\right)-\cos(\pi\sigma)\Phi\left(\frac{1}{y\alpha}\right)\right]
\end{equation}
with the parameter $\sigma$ defined as 
\begin{equation}
\sigma=\frac{1}{2}\left(\frac{k^2}{q^2}-1\right),\quad q^2=\frac{2\pi H}{\phi_0},
\end{equation}
and $\Phi(x)=\frac{2}{\sqrt{\pi}}\int^{\infty}_{x} e^{-z^2}dz$ being the complementary error function. The dimensionless parameter $\alpha$ that enters the definition of $S$ has the form 
\begin{equation}
\alpha=\frac{\sqrt{h}}{\lambda+t(2n+1)},
\end{equation}
where we have introduced dimensionless temperature $t=T/T_c$, dimensionless magnetic field $h=2eH(v_F/2\pi T_c)^2$, and scattering parameter $\lambda=1/(2\pi T_c\tau)$. 

\subsection{Benchmark for Helfand-Werthamer solution of $H_{c2}(T,\tau)$}

\begin{figure}
\includegraphics[width=9cm]{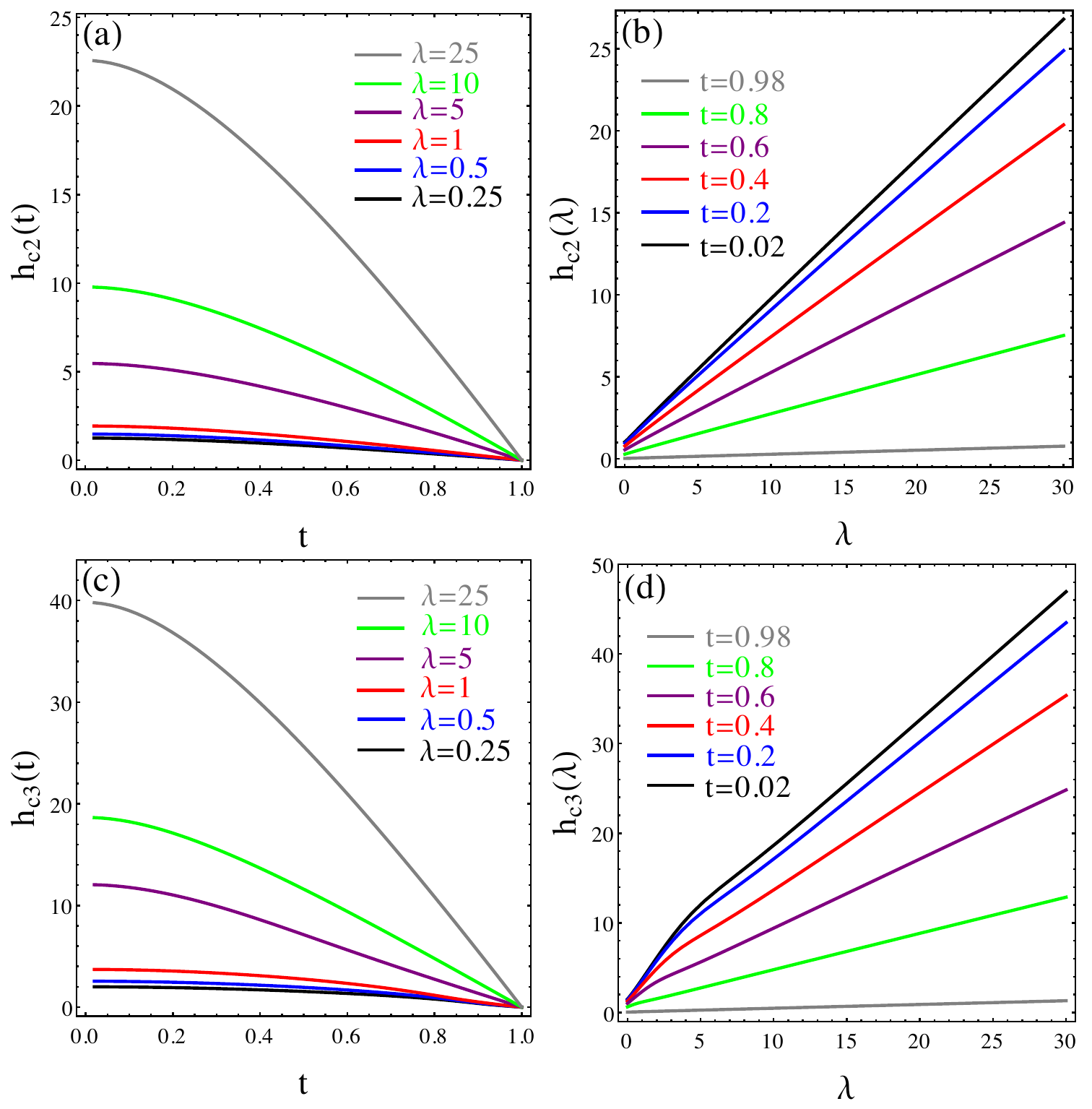} 
\caption{[Color online]: Panels (a) and (c) represent results for the critical fields $h_{c2}$ and $h_{c3}$ as a function of temperature computed from Eqs. \eqref{hc2} and \eqref{hc3} for several different values of the scattering parameter $\lambda=0.25, 0.5, 1, 5, 10, 25$ (from the lowest to the top curve respectively). Panels (b) and (d) represent the same data but plotted now as a function of scattering parameter for several different values of dimensionless temperature $t=0.02, 0.2, 0.4, 0.6, 0.8, 0.98$ (from the lowest to upper curve, respectively).}
\label{fig-Hc23}
\end{figure}

For the consistency of presentation we are going to demonstrate within this subsection that the above scheme of equations reduces to the Helfand-Werthamer solution for the $H_{c2}(T,\tau)=\phi_0/(2\pi\xi^2)$, which corresponds to $k^2/q^2=-1$, so that $\sigma=-1$. In this case, $H_{c2}$ marks the lowest eigenvalue of Eq. \eqref{Delta}. For that purpose we define $J_\sigma(\alpha)=\alpha S$ so that 
\begin{equation}
J(\alpha)\equiv J_{-1}(\alpha)=\int^{1}_{0}\frac{\sqrt{\pi}dy}{1+y^2}\left[\Phi\left(\frac{y}{\alpha}\right)+\Phi\left(\frac{1}{y\alpha}\right)\right].
\end{equation}  
We integrate by parts and get 
\begin{eqnarray}
J(\alpha)=\sqrt{\pi}\left[\frac{\pi}{2}\Phi\left(\frac{1}{\alpha}\right)+\int^{1}_{0}\frac{2dy}{\sqrt{\pi}\alpha}\arctan(y)e^{-y^2/\alpha^2}\right. \nonumber \\ \left. -\int^{1}_{0}\frac{2dy}{\sqrt{\pi}\alpha y^2}\arctan(y)e^{-1/(\alpha^2 y^2)}\right].
\end{eqnarray}
In  the first integral we rescale $x=y/\alpha$, while in the second integral we first change $y\to1/y$, use $\arctan(y)+\arctan(1/y)=\pi/2$, and also rescale to $x$ so that we obtain      
\begin{equation}
J(\alpha)=2\int^{\infty}_{0}\exp(-x^2)\arctan(\alpha x)dx. 
\end{equation}
This function appears in the Helfand-Werthamer paper Ref. [\onlinecite{HW}] [see their Eq. (25)] with exactly the same convention for $\alpha$. We bring this result now into the self-consistency condition to find 
\begin{equation}
\ln\frac{1}{t}=2\sum_{n>0}\left[\frac{1}{2n+1}-\frac{tJ(\alpha)}{\sqrt{h_{c2}}-\lambda J(\alpha)}\right],
\end{equation}
which implicitly defines $H_{c2}$ ($h_{c2}$ in the dimensionless notations), as a function of $t=T/T_c$ for any value of $\lambda$. For the numerical computation it is advantageous to rewrite this equation in the form that does not contain spurious logarithmic divergence. To this end, we introduce $\widetilde{J}(\alpha)=[\alpha-J(\alpha)]/\alpha^3$ so that $\lim_{n\to\infty}\widetilde{J}(\alpha)=1/3$ and $\widetilde{J}(\alpha)>0$. In these notations  
\begin{equation} \label{hc2}
\ln\frac{1}{t}=\sum_{n \ge 0}
\frac{2 h^{3/2}_{c2} \widetilde{J}(\alpha)}{(2n+1)[\lambda+t(2n+1)]^2[\sqrt{h_{c2}} - \lambda J(\alpha)]} .
\end{equation} 
This equation was solved numerically and the results for $h_{c2}(t,\lambda)$ are presented in the Figs. \ref{fig-Hc23}(a-b). For fixed values of the disorder parameter $\lambda$, the critical field is monotonically growing function with lowered temperature $t$. In the strongly ballistic case $T_c\tau\gg1$, the temperature dependence of $h_{c2}$ is insignificant. However, $h_{c2}$ is a very sensitive function of the disorder parameter and scales linearly with $\lambda$ with the temperature-dependent slope. This dependence can be simply understood. In the ballistic case, the coherence length scales as $\xi_b\sim v_F/T_c$, whereas in the diffusive case it is $\xi_d\sim \sqrt{D/T_c}$, where $D=v^2_F\tau/3$ is the diffusion coefficient. Since the critical field is inversely proportional to the square of the coherence length, then consequently the ratio of critical fields in the two limiting cases is $h^d_{c2}/h^b_{c2}\sim(\xi_b/\xi_d)^2\sim 1/(T_c\tau)\propto\lambda$. These results are consistent with HW [\onlinecite{HW}].    

\subsection{Temperature and purity dependence of $H_{c3}(T,\tau)$}

\begin{figure}
\includegraphics[width=9cm]{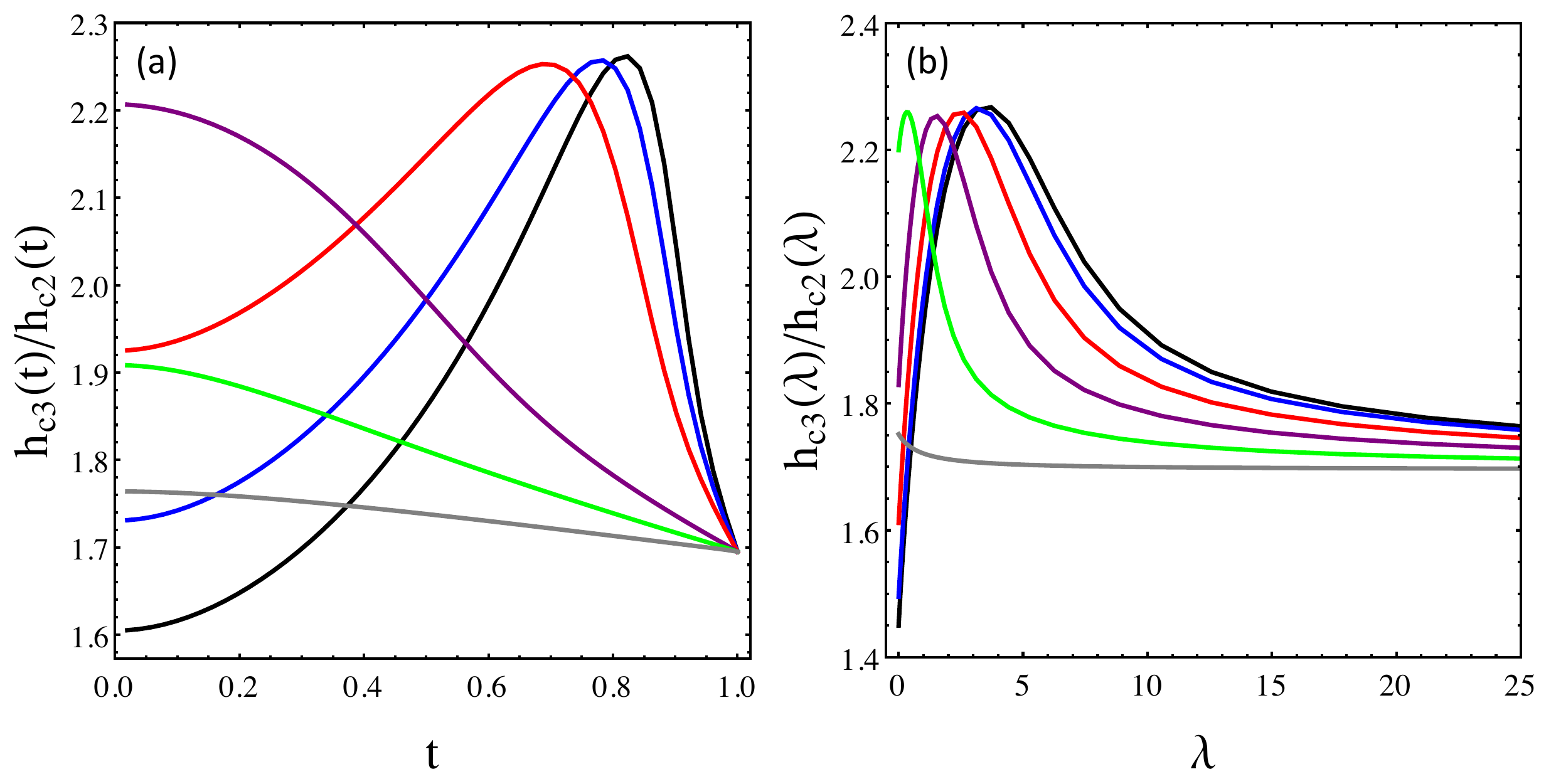} 
\caption{[Color online]: Panel (a) shows the ratio between third and second critical fields as a function of temperature for the same choice of the scattering parameters as in Fig. \ref{fig-Hc23} (a,c), while panel (b) shows the same ratio but plotted as a function of the scattering parameter for different values of temperature in conjunction with Fig. \ref{fig-Hc23}(b,d).}
\label{fig-Hc3-Hc2}
\end{figure}

Equation \eqref{GL} formally coincides with the linearized GL equation. In terms of the SJdG solution for $H_{c3}$ near $T_c$ with the conventional boundary condition $F'=0$ at the plane surface of a half-space one finds $(q\xi)^2=1.695$. Hence, for the field and temperature dependence of $\xi$, we therefore have the same result where $\xi(T,H)$ should be evaluated now with the help of self-consistency equation \eqref{SC-Eq} of the theory. Within our formalism one has to set $\sigma=-\frac{1}{2}\left(1+1/(q\xi)^2\right)=-0.795$ in the expression \eqref{S} for $S$ and solve \eqref{SC-Eq}. In doing so, we introduce 
\begin{eqnarray}
J_\sigma(\alpha)&=&\sqrt{\pi}\alpha\int^{\infty}_{0}dx\frac{(1+\alpha^2x^2)^\sigma}{(1-\alpha^2x^2)^{\sigma+1}}
\nonumber\\ 
&&[\theta(1-\alpha x)-\theta(\alpha x-1)\cos(\pi\sigma)]\Phi(x),
\end{eqnarray} 
and also $\widetilde{J}_\sigma=[\alpha-J_\sigma(\alpha)]/\alpha^3$. In terms of these functions we find from \eqref{S} the desired equation for the critical field of superconductivity nucleation in the form 
\begin{equation} \label{hc3}
\ln\frac{1}{t}=\sum_{n \ge 0}
\frac{2 h^{3/2}_{c3} \widetilde{J}_\sigma(\alpha)}{(2n+1)[\lambda+t(2n+1)]^2[\sqrt{h_{c3}} - \lambda J_\sigma(\alpha)]},
\end{equation} 
which is structurally analogous to Eq. \eqref{hc2}. This equation also admits straightforward numerical solution that we present in Fig. \ref{fig-Hc23}(c-d). The behavior of $h_{c3}$ as a function of either $t$ or $\lambda$ is very similar to that of $h_{c2}$. The most significant differences are revealed when we plot the ratio of two fields. Interestingly, the $H_{c3}/H_{c2}$ is in fact a nonmonotonic function of temperature for a parameter range of the scattering coefficient $\lambda\sim1$ that corresponds to a crossover region from ballistic-to-diffusive scattering in a bulk of a superconductor [see Fig. \ref{fig-Hc3-Hc2}(a)]. As the temperature approaches $T_c$, the ratio $H_{c3}/H_{c2}$ tends to SJdG limit in accordance with the GL theory.
In general, for a large range of temperatures this is not the case. The maximum occurs at about $\mathrm{max}\{H_{c3}/H_{c2}\}\approx2.25$. For strongly disordered case, the temperature dependence of the ratio is insignificant and the magnitude is close to SJdG limit for all temperatures. The same nonmonotonicity is revealed for the $H_{c3}/H_{c2}$ ratio when plotted versus scattering parameter $\lambda$ for different temperatures [see Fig. \ref{fig-Hc3-Hc2}(b)]. Again, in the limit when $\lambda\gg1$ all lines tend to converge to SJdG result for any temperature.   

\section{Boundary effect on surface superconductivity}\label{Sec:Boundary-SSC}

Surface superconductivity is a robust and quite pronounced phenomenon as observed in polished clean material structures and films. One may naturally expect that superconductivity of the surface layer will be extremely sensitive to the quality of the surface and in general will be fragile for rough surfaces mediating substantial surface scattering. However, this question has not received sufficient theoretical attention. We address this problem in the simplest situation assuming that bulk of the superconductor is ballistic and scattering takes place only at the boundary. For that purpose we also concentrate in this section only on the case of zero temperature. The convenient starting point is the formalism of Abrikosov [\onlinecite{Abrikosov}] and Hu-Korenman [\onlinecite{Hu-Korenman}] who studied the case of specular reflection. They showed that linearized Gor'kov's equation for the self-consistent order parameter in the one-dimensional geometry reduces to the following linear integral problem [\onlinecite{Abrikosov,Hu-Korenman}]       
\begin{align}\label{IntEq}
&L\Delta(x)=K_0(x)\Delta(x)+ \nonumber \\ 
& \int^{\infty}_{0}\!\!K_1(x,x')[\Delta(x)-\Delta(x')]dx' 
-\int^{\infty}_{0}\!\!K_2(x,x')\Delta(x')dx',
\end{align}
where $L=\ln(\mathbb{C} T_cl_H/v_F)$ and $\mathbb{C}=\pi e^{-\gamma_E/2}$ with $\gamma_E\approx0.577$ being the Euler-Mascheroni constant. It is assumed that the boundary is located at $x=0$ and the superconductor occupies the volume $x>0$. All lengths were scaled by magnetic length $l_H=\sqrt{1/eH}$. The integral kernels are given by the following expressions:
\begin{align}
&K_0=\int^{|x-x_0|}_{0}e^{-(x-x_0)^2+x'^2}x'\ln\left[\frac{|x-x_0|+x'}{|x-x_0|-x'}\right]dx'+\nonumber \\ 
&+\frac{1}{2}\int^{\infty}_{0}\frac{e^{-|x(x-2x_0)-x'(x'+2x_0)|}}{x+x'}dx', \label{K0} \\
&K_1=\frac{1}{2|x-x'|}e^{-|x(x-2x_0)-x'(x'-2x_0)|},\label{K1}\\
&K_2=\frac{1}{2(x+x')}e^{-|x(x-2x_0)+x'(x'-2x_0)|}\label{K2},
\end{align} 
where the parameter $x_0$ in these kernels marks the center of the nucleation. Equation \eqref{IntEq} can be considered as the eigenvalue problem for the linear integral operator acting on $\Delta$, where the eigenvalue itself is a function of $x_0$. The lowest value of the eigenvalue achieved for the optimal $x_0$ defines the highest possible field for which finite solution exists. While the problem at hand of finding that eigenvalue is conceptually simple, it is technically difficult as the eigenvalue equation has no scale and as such all parameters are of the order of unity. Also owing to the complex structure of the kernels there is no obvious way to solve the equation \eqref{IntEq}. For that reason, Hu-Korenman [\onlinecite{Hu-Korenman}] relied on the variational ansatz. This computational approach is very reasonable since as shown by Helfand and Werthamer [\onlinecite{HW}], a similar variational calculation for the problem of $H_{c2}$ as carried by Gor'kov [\onlinecite{Gorkov}] yields an exact result. We will briefly recap these results in the following section. Next, we will provide generalization of Eq. \eqref{IntEq} to the case of diffusive scattering. We solve the corresponding eigenvalue exactly by numerical diagonalization and also suggest a modified variational wave function that captures the main results with the greater precision.     

\subsection{Benchmark for Gor'kov's solution of $H_{c2}$}

To gain the technical grasp of the problem. it is useful to recapture Gor'kov's integral equation for $H_{c2}$ from Eq. \eqref{IntEq}. In the $H_{c2}$ problem vortex nucleation occurs in the bulk of a superconductor, in other words its center is at infinity with respect to the boundary. To access this limit, one has to shift variables by  introducing $x\to x-x_0$ and $x'\to x'-x_0$, and then take limit of $x_0\to\infty$. In this case, the kernel $K_2$ vanishes as is clear from Eq. \eqref{K2}, as well as the second term in the kernel $K_0$ because of its exponential dependence on $x_0$.  As a result, Eq. \eqref{IntEq} simplifies to [\onlinecite{Gorkov}]
\begin{equation}\label{IntEq-Gorkov}
L\Delta(x)=K_0(x)\Delta(x)+\int^{+\infty}_{-\infty}\frac{e^{-|x^2-x'^2|}}{2|x-x'|}[\Delta(x)-\Delta(x')]dx'
\end{equation} 
with $K_0\to e^{-x^2}\int^{|x|}_{0}x'e^{x'^2}\ln\frac{|x|+x'}{|x|-x'}dx'$. As the next step, we adopt a simple Gaussian trial wave function with a single parameter $a$:
\begin{equation}\label{Delta-Trial1}
\Delta(x)=\exp(-ax^2).
\end{equation}
Multiplying now Eq. \eqref{IntEq-Gorkov} by $\Delta(x)$ from Eq. \eqref{Delta-Trial1} and integrating over $x$ we obtain $L$ as a function of $a$. All integrals are carried out in Appendix \ref{appendix-integrals}  and can be completed analytically by passing to hyperbolic coordinates. We thus find from Eq. \eqref{IntEq-Gorkov}  
\begin{equation}\label{Hc2-Gorkov}
\ln\left(\frac{\mathbb{C}T_c}{v_F}\sqrt{\frac{1}{eH}}\right)=\frac{1}{2}\ln\left[\frac{(1+a)^2}{2a}\right].
\end{equation}
The right-hand side of this equation has an extremum at $a=1$ that defines the upper critical field at zero temperature of a clean superconductor $H_{c2}=\mathbb{C}^2T^2_c/2ev^2_F$. 

This method of finding $H_{c2}$ was generalized by Shapoval to compute critical fields in the case of superconducting films with rough surfaces [\onlinecite{Shapoval}]. Next, we employ this approach to the problem of $H_{c3}$.   

\begin{figure}
\includegraphics[width=0.22\textwidth]{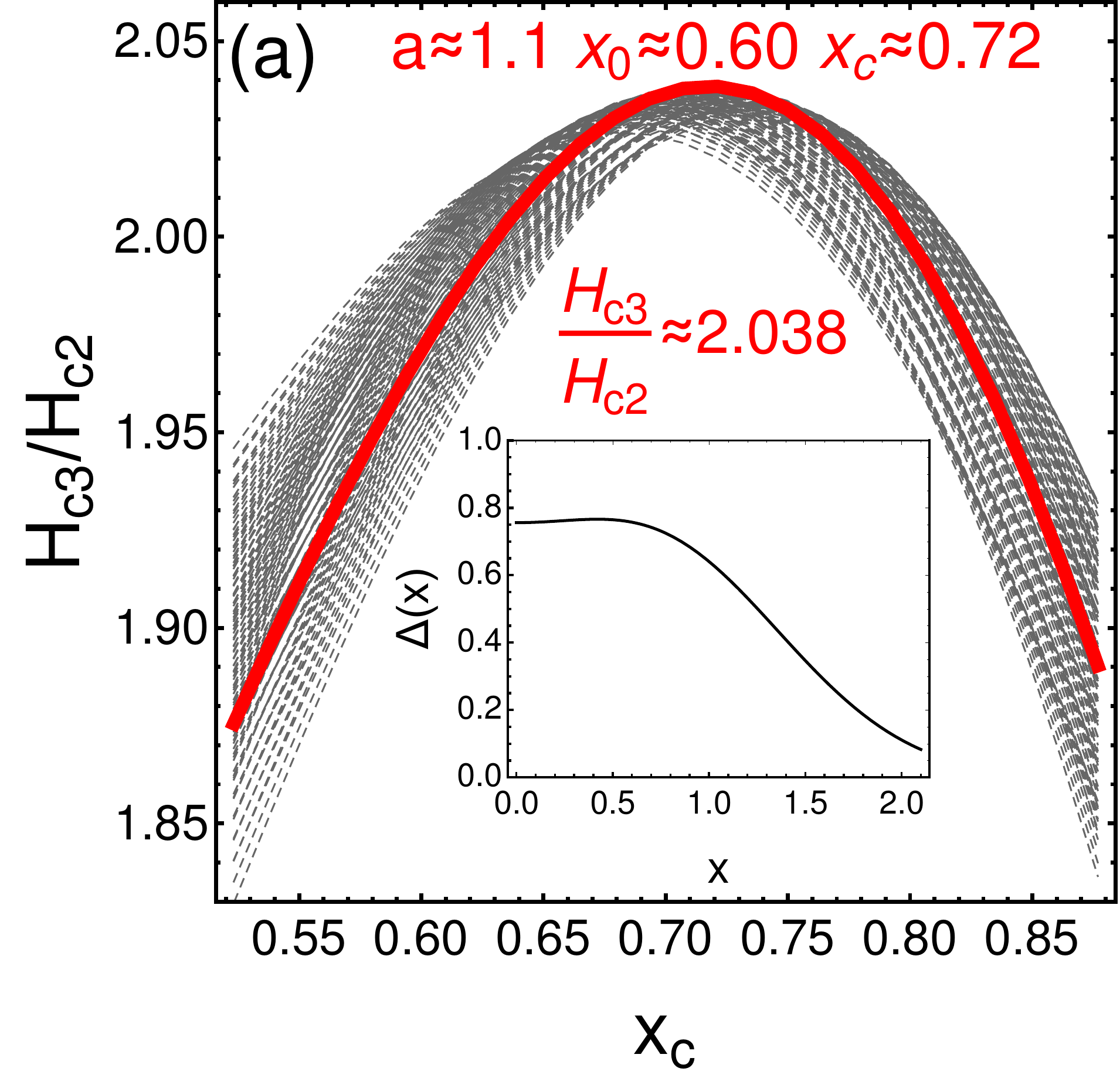}
\includegraphics[width=0.235\textwidth]{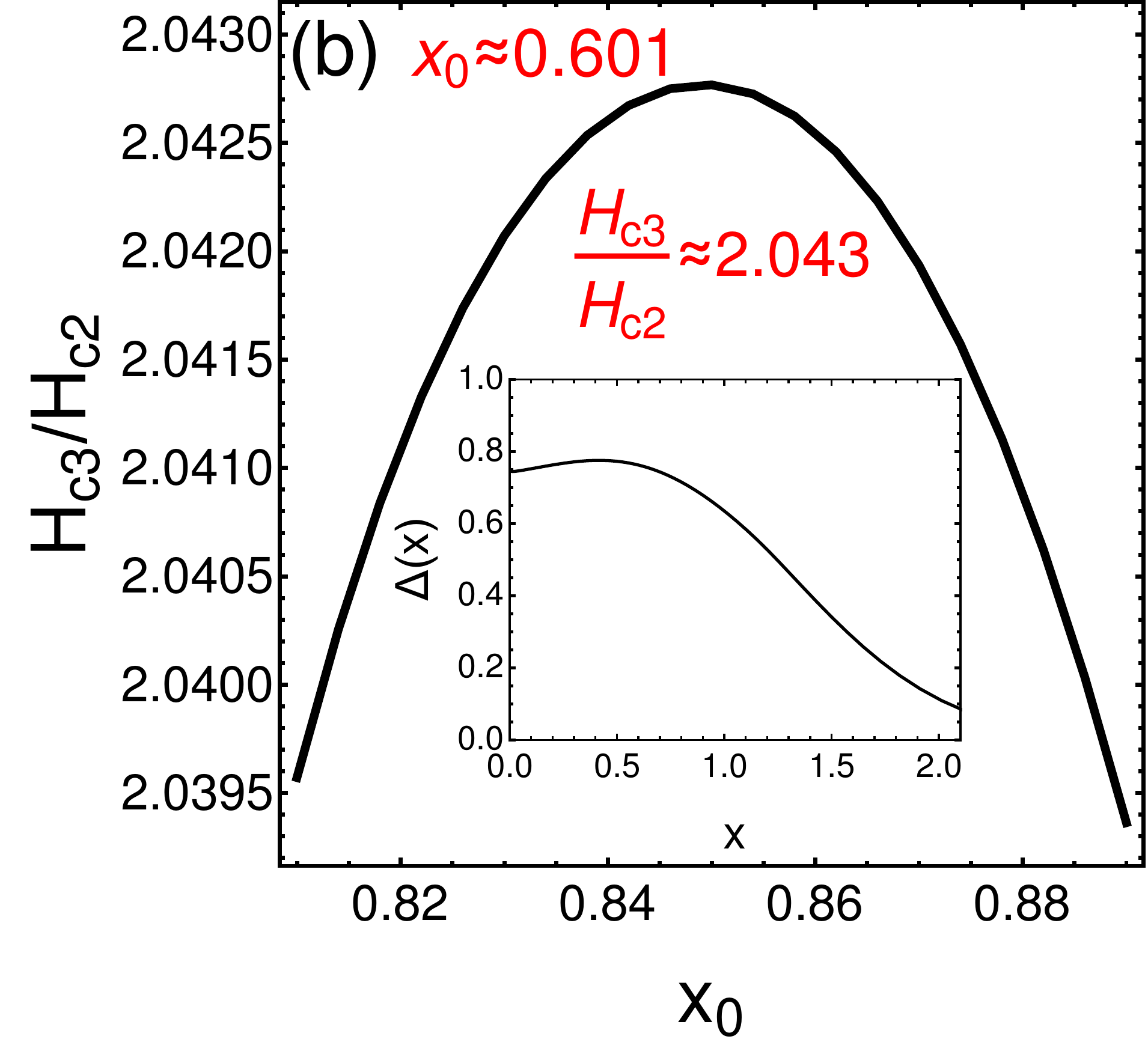}
\caption{[Color online]: Results for the $H_{c3}/H_{c2}$ ratio at zero temperature for the case of diffusive scattering from the boundary and assumed ballistic limit in the bulk of a superconductor. Panel (a) displays results of the variational computation which is based on the trial wave function from Eq. \eqref{Delta-Trial2}. Optimal fitting parameters $a,x_0,x_c$ are listed on the plot whereas inset shows the profile on the nucleated superconducting order parameter near the surface. Panel (b) shows results for the same quantities but obtained via unbiased exact solution of the integral eigenvalue equation \eqref{IntEq}.} 
\label{fig-Hc3-Diffusive}
\end{figure}

\subsection{Robustness of $H_{c3}$ for diffusively reflecting boundary} 

In order to address the effect of surface scattering on critical fields of superconductors, one may benefit from the early work by Fuchs on the electronic conductivity in thin metallic layers [\onlinecite{Fuchs}]. In the Fuchs model, the roughness of the edge is modeled by treating the reflection angle of each semiclassical trajectory that collides with the side edges of the sample as an independent random variable. This was a basis for Shapoval's solution of the linearized gap equation in superconducting films. Repeating the same calculations as in Refs. [\onlinecite{Abrikosov,Hu-Korenman,Shapoval}] but for the case of a single boundary in the context of surface superconductivity and averaging gap equation over the angle of scattering off the boundary, we arrive at an eigenvalue problem for $H_{c3}$ which is structurally identical to Eq. \eqref{IntEq}. The only difference is that in the third term the kernel $K_2(x,x')\to K_3(x,x')$ has to be replaced by a different expression
\begin{eqnarray}\label{K3}
K_{3}(x,x')=\iint^{\infty}_{0}\frac{k(u,x)k(u',x')dudu'}{x\sqrt{1+u^2}+x'\sqrt{1+u'^2}}, \\ 
k(u,x)=\frac{u}{(1+u^2)^{3/2}}J_0\left(ux(x-2x_0)\right),
\end{eqnarray} 
where $J_0$ is the Bessel function and otherwise the notations are identical to that in Eq. \eqref{IntEq}. If not for a $u$-dependent factors in the denominator of Eq. \eqref{K3}, that comes from angular averaging of random reflections, the $u$-integrals would separate and by virtue of an integral identity $\int^{\infty}_{0}\frac{uJ_0(au)du}{(1+u^2)^{3/2}}=e^{-a}$, one would recover Eq. \eqref{K2} from \eqref{K3} valid for the specular boundary condition. 

To solve Eq. \eqref{IntEq} for the diffusive scattering case it is advantageous to separate the kernel 
$K_3$ from Eq. \eqref{K3} into singular-$K^s_3$ and regular-$K^r_3$ parts. First, observe that as $\{x,x'\}\to0$ the asymptotic form of Eq. \eqref{K3} is as follows: $K_3(x,x')\to f(x'/x)/x+f(x/x')/x'=F(x'/x)/(x+x')$ where $f(z)=\frac{1}{3}[1-z\ln(1+z^{-1})]$ that is analytical and $0\leq f(z)\leq 1/3$ for all $z\geq0$, and $F(z)=(1+z)f(z)+(1+z^{-1})f(z^{-1})$ that is also analytical for all $z\geq0$ where $\frac{4}{3}(1-\ln2)\leq F(z)\leq 1/2$. These observations suggest the following decomposition 
\begin{align}
&K_3(x,x')=K^s_3(x,x')+K^r_3(x,x'), \\ 
&K^s_3(x,x')=\frac{F(x'/x)}{(x+x')}e^{-|x(x-2x_0)+x'(x'-2x_0)|},\\
&K^r_3(x,x')=\iint^{\infty}_{0}\frac{dudu'}{x\sqrt{1+u^2}+x'\sqrt{1+u'^2}}\times\nonumber \\ 
&[k(u,x)k(u',x')-e^{-|x(x-2x_0)+x'(x'-2x_0)|}k(u,0)k(u',0)]
\end{align}
that enables to reformulate Eq. \eqref{IntEq} that is more computationally stable for the numerical analysis. The algorithm that we used to numerically solve Eq. \eqref{IntEq} by an exact diagonalizaton of an eigenvalue problem for $K_2\to K_3$ diffusive case is described in the Appendix \ref{appendix-algorithm}. We have also used an improved variational approach with the trial wave function of the form 
\begin{equation}\label{Delta-Trial2}
\Delta(x)=\frac{1}{2}\left[e^{-a(x-x_c)^2}+e^{-a(x+x_c)^2}\right]
\end{equation}
for the comparison of two computations. In contrast to Eq. \eqref{Delta-Trial1} that provides a solution localized at the boundary, the form of Eq. \eqref{Delta-Trial2} incorporates an additional parameter $x_c$ that may tune the location of the localization center. 

The results of computation are shown in Fig. \ref{fig-Hc3-Diffusive}. The trial solution from Eq. \eqref{Delta-Trial2} gives the optimal estimation $H_{c3}/H_{c2}\approx2.038$ for $a\approx1.1$, $x_c\approx 0.72$, and $x_0\approx0.60$, see Fig. \ref{fig-Hc3-Diffusive}(a). The order parameter near the surface is peaked at $x^*\approx0.537$. An exact numerical solution unbiased from extra fitting parameters is depicted in Fig. \ref{fig-Hc3-Diffusive}(b) and gives $H_{c3}/H_{c2}\approx2.043$ for $x_0\approx0.601$ with the peak of nucleated order parameter at $x^*\approx0.415$. The main conclusion we draw from this analysis is that surface imperfections and disorder that leads to diffusive boundary scattering does not suppress the magnitude of the critical field for surface superconductivity nucleation. Perhaps surprisingly, we find that this field is above the SJdG limit.    

\section{Ginzburg-Landau formalism for $H_{c3}$}\label{Sec:Hc3-Disc}

In this section we turn our attention to the problem of surface superconductivity in the systems having cylindrical geometry. In contrast to the previous sections, our motivation here is not so much to reveal the dependence of $H_{c3}$ on various parameters relevant to this geometry and type of scattering, but rather to explore interesting possible consequences of surface effects for the vortex nucleation in superconducting mesoscopic disks [\onlinecite{Kanada,Roditchev}] and magnetoresistance measurements in quantum wires as motivated by numerous experiments [\onlinecite{Dynes,Kwok,Shahar,Markovic}].  For that purpose we choose to work with GL formalism whose simplicity often compensates for the lack of microscopic rigor when it comes to behavior of various observables away from GL region.   

We begin with the GL equation for the order parameter wave function $\Psi$:
\begin{equation}\label{GL-Eq}
\frac{1}{2m}\left(-i\bm{\nabla}-2e\bm{A}\right)^2\Psi+\alpha\Psi+\beta\Psi|\Psi|^2=0,
\end{equation} 
and corresponding complementary expression for the current 
\begin{equation}\label{GL-J}
\bm{J}=\frac{ie}{2m}[\Psi\bm{\nabla}\Psi^*-\Psi^*\bm{\nabla}\Psi]-\frac{2e^2}{m}\bm{A}|\Psi|^2. 
\end{equation}
Focusing on magnetic fields above $H_{c3}$ in the normal state, it is sufficient to linearize the GL equation, thus, we will drop the $\sim\beta\Psi|\Psi|^2$ term from Eq. \eqref{GL-Eq}. In this section we follow the notations of Ref.~[\onlinecite{Saint-James}]. We consider the solid cylindrical superconductor symmetric with respect to the $z$ axis. For a uniform and constant magnetic field $\bm{H}$ parallel to the $z$-axis it is convenient to choose vector potential in the Coulomb gauge $\bm{A}=\frac{1}{2}\bm{H}\times\bm{r}$. In cylindrical coordinates we have for the Laplacian $\nabla^2\Psi=\rho^{-1}\partial_\rho(\rho\partial_\rho\Psi)+\rho^{-2}\partial^2_\varphi\Psi+\partial^2_z\Psi$ and for the gradient operator $\bm{\nabla}=\hat{\rho}\partial_\rho+\hat{\varphi}\rho^{-1}\partial_\varphi+\hat{z}\partial_z$. We look for the solution of linearized Eq. \eqref{GL-Eq} in the form 
\begin{equation}\label{GL-Psi}
\Psi(\phi,\rho,z)=e^{ik_zz}e^{im_L\varphi}f(\rho)
\end{equation} 
with the angular momentum $m_L$. The lowest eigenvalue results from the solution with $k_z=0$ so that we have 
\begin{equation}\label{GL-A}
-\left[\frac{1}{x}\partial_x(x\partial_xf)-\frac{m^2_L}{x^2}f\right]-hm_Lf+\frac{h^2}{4}x^2f=f.
\end{equation} 
We arrived at Eq.~\eqref{GL-A} by dividing the GL equation, \eqref{GL-Eq} by the characteristic energy, $|\alpha|$.
In equation \eqref{GL-A} we introduced the dimensionless coordinate, $x=\rho/\xi_{GL}$ normalized by temperature dependent GL coherence length $\xi^{2}_{GL}=1/(2m|\alpha|)$, and dimensionless field, $h=H/H_{c2}$ normalized by the upper critical field in GL theory $H_{c2}=m|\alpha|/e$. We are after the maximal field $h=h_{c3}$ such that Eq. \eqref{GL-A} has a solution with the boundary condition
\begin{equation}\label{BC}
\partial_x f(x)|_{x=x_R}=0, \quad x_R=R/\xi_{GL}, 
\end{equation}  
where $R$ is the radius of the cylinder. Note that GL theory requires only that the radius $R$ is larger than the size of the Cooper pair which is of the order of $\sim v_F/T_c$ or coherence length at zero temperature, while the radius $R$ can be both smaller and larger than the GL coherence length $\xi_{GL}$, so that $x_R$ can in principle take any value.    

We are focusing on the fields $h > 1$. In the limit of finite but large radius of the cylinder, $R \gg \xi_{GL}$ we expect to recover the result for $h_{c3} = 1.695$ for a half-space. However, we expect to find $h_{c3}$ in excess of this value the radius of the cylinder becomes comparable with the the GL coherence length.

\begin{figure}
\includegraphics[width=0.21\textwidth]{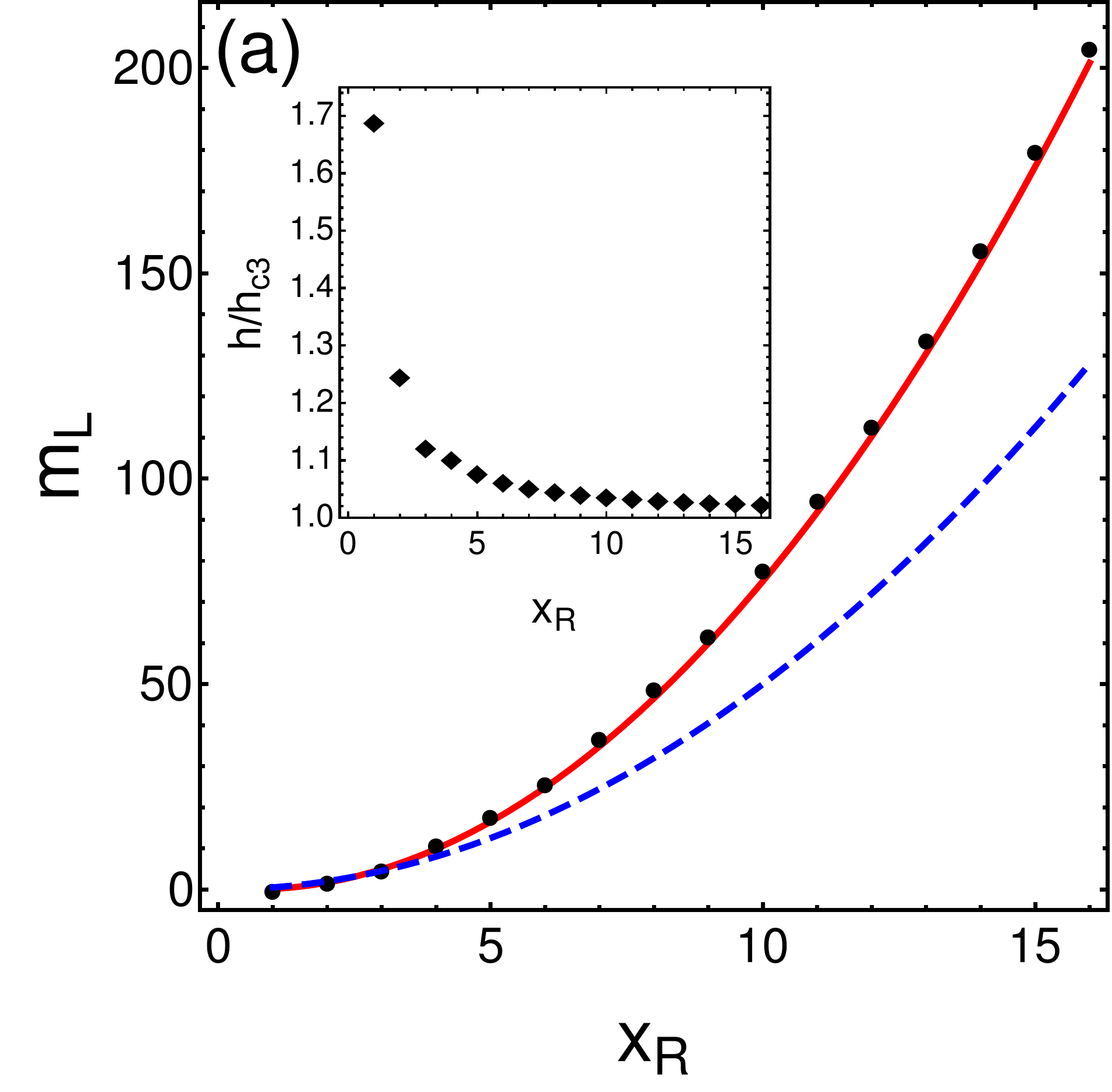}
\includegraphics[width=0.25\textwidth]{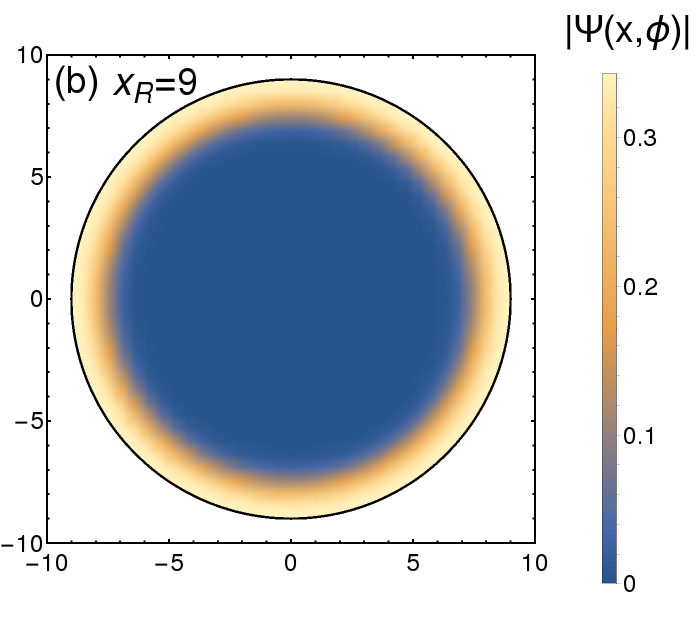}
\caption{[Color online]: On the panel (a) the points are obtained by numerically solving $[\partial_x f](x_R,m_L,h) = 0$ with $f$ taken from Eq. \eqref{GL-f} for few values of the radius of the disk. The agreement between the points and the red line just confirms that we reproduce the limit of the semi-infinite superconductor when $x_R \gg 1$ [this limit is reached starting from $x_R =5$ i.e. for $R = 5 \xi_{GL}(T)$]. The blue dashed line is the relation $m_L=x^2_R/2$ obtained for the points giving the $h_{c2}$ of the semi-infinite case. The inset shows the ratio between $h_{c3}$ in the disk and $h_{c3}$ of the semi-infinite planar geometry for several values of $x_R$. The ratio converges to unity once $x_R\gg1$. Panel (b) shows the spatial profile of the localized surface superconductivity as described by the GL wave function $\Psi$. } 
\label{fig-Hc3-GL}
\end{figure}

To understand the crossover to the problem of a semi-infinite superconductor that occurs at large $R \gg \xi_{GL}$, we write \eqref{GL-A} in the form
\begin{equation}\label{GL_re}
- \frac{1 }{ x} \partial_x \left( x \partial_x f \right)  + V(x,m_L,h) f   = f,
\end{equation}
where the effective potential 
\begin{equation}\label{V}
V(x,m_L,h) = \frac{m_L^2}{ x^2 }   - h m_L  + \frac{h^2}{4} x^2 = \left( \frac{m_L}{x} - \frac{h}{2} x \right)^2,
\end{equation}
has a minimum at 
\begin{equation}\label{x-min}
x_{\min} = \sqrt{ 2 m_L / h}\, ,
\end{equation}
which is also a scale of variation of the potential, Eq.~\eqref{V}.

Assuming for now that $m_L \gg 1$ and $h \simeq 1$, we have $x_{\min} \gg 1$. Subject to further confirmations, the lowest-energy solutions of \eqref{GL_re} are localized near the minimum of the potential, \eqref{V} on a scale, $\Delta x \ll x_{\min}$, as given by \eqref{x-min}. Then, we may keep only the first nonvanishing term in the expansion of the potential \eqref{V} near $x_{\min}$ which replaces Eqs.~\eqref{GL_re} and \eqref{V} with the effective one-dimensional problem,
\begin{equation}\label{GL-1D}
-  \partial_x^2  f + h^2 ( x - x_{\min})^2 f   = f.
\end{equation}
The parameter $x_{\min}$ in Eq.~\eqref{GL-1D} plays the role of the guiding center in the semi-infinite superconductor. The characteristic spatial extent of its solutions, $\Delta x \sim 1/\sqrt{h}$ is of order one since $h \simeq 1$. Therefore, Eq.~\eqref{GL-1D} is a valid approximation of Eq.~\eqref{GL_re} provided only that $x_{\min} \gg 1$ which is realizable in large cylinders, $R\gg \xi_{GL}$. In the range of angular momenta, $m_L$ such as $ 1 \ll  x_{\min} \lesssim R/\xi_{GL} - 1$ the boundary condition, \eqref{BC} is inessential. 
Hence, for large cylinders the dense set of circularly symmetric degenerate states with the eigenvalue $h$ closely resembles the solutions of the semi-infinite space problem. Clearly, the vortices of the bulk are their linear superpositions up to a gauge transformation in case different gauges have been used for the two problems. The vortices then according to Eq.~\eqref{GL_re}  form at the bulk upper critical field $h =1$,  consistent with our present notation for $h$ as the ratio of the field, $H$ to $H_{c_2}$. Apart from an obvious lower bound, $h_{c3} \geq h_{c2}$, the above argument links the large cylinder problem to the problem of the semi-infinite superconductor. This correspondence holds for $m_L$ that are not necessarily large.
In fact, for $m_L=0$, the original Eq.~\eqref{GL-A}  takes the form, $- x^{-1} \partial_x \left( x \partial_x f \right)  + (h/2)^2 x^2 f = f$. The solutions of this equation are isotropic wave functions of the two-dimensional harmonic oscillator with the same lowest eigenvalue $h$ and Gaussian wave function localized at the origin. 

\begin{figure}
\includegraphics[width=0.4\textwidth]{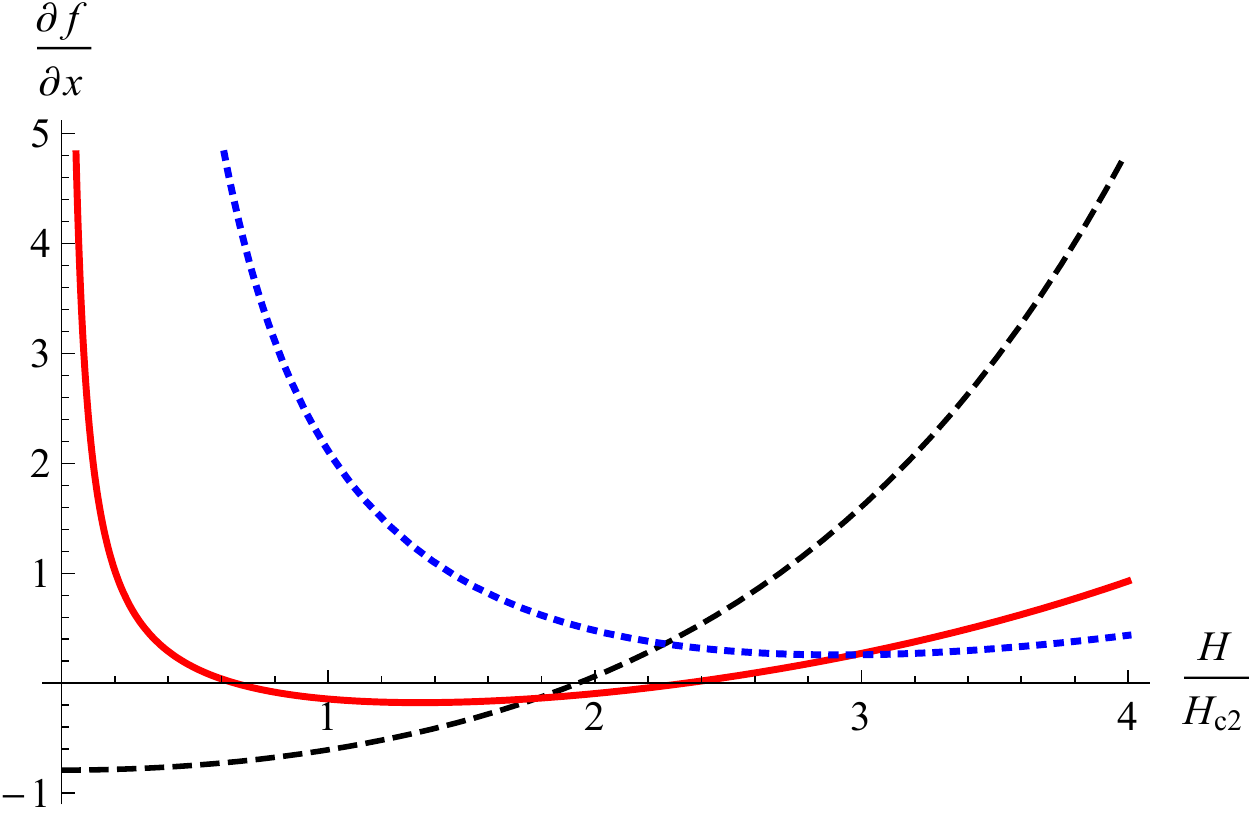}
\caption{[Color online]: Here we consider the disk with the radius, $R =(3/2) \xi_{GL}$ as in Ref. [\onlinecite{Roditchev}], and plot the derivative $[\partial_x f](3/2,m_L,h)$ where the function $f$ is given by Eq. \eqref{GL-f} for $m_L = 0$ (black, dashed line), $m_L = 1$ (red, solid line), $m_L = 2$ (blue, dotted line). We see that the $m_L = 1$ curve gives the highest field yielding the zero of the derivative  $[\partial_x f](3/2,m_L,h)$. This agrees with the observations of Ref. [\onlinecite{Roditchev}]. We further notice that for this disk, $H_{c3}(x_R=1.5) \approx 2.35 H_{c2}$ which is about $1.4$ times higher than the bulk $H_{c3}$, which is in agreement with the general dependence of the $H_{c3}$ on the radius of the disk [see also inset in Fig. \ref{fig-Hc3-GL}(a)].} 
\label{fig-dfdx}
\end{figure}

The exact solutions of equation \eqref{GL-A} for $m_L=0,1,2,\ldots$ that are regular at the origin are expressed through the hypergeometric function,
\begin{equation}\label{GL-f}
f(x,m_L,h)\!=\! x^{m_L} e^{-hx^2/4} \phantom{}_1F_1\!\left(\frac{h-1}{2h},m_L+1,\frac{hx^2}{2}\!\right)\, ,
\end{equation} 
and for $m_L = -1,-2,\ldots$
\begin{equation}\label{GL-fA}
f(x,m_L,h) \!=\! x^{-m_L} e^{-hx^2/4} \phantom{}_1F_1\!\left( \frac{ h - 1}{ 2 h } - m_L,1- m_L  , \frac{h x^2}{2}\! \right).
\end{equation}
This can be verified by the direct substitution of Eqs.~\eqref{GL-f}, \eqref{GL-fA} into Eq.~\eqref{GL-A}.
To determine the $h_{c_3}$ one has to find the $h$ solving the equation $f_x(R/\xi_{GL},m_L,h)=0$, where $f_x(x,m_L,h) \equiv  \partial_x f(x,m_L,h)$ for different values of $m_L$.
The actual $h_{c_3}$ is the maximal among all these solutions. 
We have verified that the solutions with negative $m_L$, [\eqref{GL-fA}] produce fields lower than the solutions with non-negative $m_L$ [\eqref{GL-f}] and therefore solutions \eqref{GL-fA} are discarded for the present discussion. 

We have checked that in the limit of large cylinder, $R \gg \xi_{GL}$ the SJdG result, $h_{c3} = 1.695$ is recovered. In the SJdG solution the guiding center is located a distance $\delta R = \mu^2 \xi_{GL}$ from the boundary which translates into the dimensionless distance $x_0 =  \delta R/ \xi_{GL}^2 = \mu^2 = 0.59$. 
This according to Eq. \eqref{x-min} implies that the relationship between the optimal angular momentum and the cylinder's radius,  
\begin{equation}\label{GL-hc3-cond}
\sqrt{ 2 m_{L} / h_{c3} } = x_R - x_0
\end{equation}
has to be satisfied with $h_{c3} = 1.695$ at $x_R \gg 1$.
Alternatively, for the optimal angular momentum,
\begin{equation}\label{m-opt}
m_{L} \approx 0.85 (x_R - 0.59)^2.
\end{equation}
We have found numerically the pairs of $(m_L,h)$ such that $[\partial_x f](x_R,m_L,h) = 0$. 
Then, among all possible such pairs we could locate the $m_L$ that gives the highest $h$, and checked that at sufficiently large $x_R$ the relation \eqref{m-opt} is satisfied, [see Fig. \ref{fig-Hc3-GL}(a)].
The spatial profile of the GL wave function Eq. \eqref{GL-Psi} corresponding to an exact solution Eq. \eqref{GL-f} is shown in Fig. \ref{fig-Hc3-GL}(b) for $x_R=9$ which reveals clear superconductivity localization at the edge of the disk. For cylinders with radii less than $5 \xi_{GL}$ the highest $h_{c_3}$ exceeds significantly the SJdG value for the semi-infinite superconductor, [see inset in Fig. \ref{fig-Hc3-GL}(a)]. Within the present approach at small $x_R$ the critical field can be arbitrarily large. We reiterate, however that the current approach applies to cylinders of the sufficiently large radius, $R \gtrsim v_F/T_c$. For smaller radii, the electromagnetic response of the superconductor is essentially nonlocal, which is beyond the local GL formulation.

The formation of the states with $m_L = 1$ was reported in the Pb nano-islands of diameter about $3 \xi_{GL}$. The $m_L=1$ forms at sufficiently high perpendicular magnetic field and retains its character up to a complete suppression of the superconductivity.  We argue that the formation of $m_L=1$ state as reported in Ref. [\onlinecite{Roditchev}] based on the scanning tunneling spectroscopy (STS) is in agreement with the present general formulation. In our terms, the superconducting to normal phase transition occurs at the field $H_{c3}$ and in fact the superconductivity is reported to be suppressed at the center of the island and gets enhanced at its edges, in agreement with the general picture of the surface superconductivity. 
To confirm the above correspondence with the STS data we have plotted in Fig. \ref{fig-dfdx} the derivative, $[\partial_x f](x_R,m_L,h)$ for $x_R = 1.5$ as in Ref. [\onlinecite{Roditchev}] for $m_L = 0$, $m_L = 1$ and $m_L = 2$ to show that only for $m_L=0$ and $m_L=1$ this derivative vanishes at a finite field.
The highest field at which $[\partial_x f](x_R,m_L,h)=0$ is obtained for $m_L=1$ and is equal to $h_{c_{3}} \approx 2.35$ which is $1.4$ times larger than the SJdG value. This results agrees with the experiment reporting the onset of the superconductivity in the single vortex state with $m_L = 1$, [\onlinecite{Roditchev}].
The increase of the $H_{c_3}$ with decreasing radius of the cylinder is a general trend. 
As indicated in the inset of Fig. \ref{fig-Hc3-GL}(a) at even smaller cylinder of a radius $R = \xi_{GL}$, $h_{c_{3}}$ is still larger than SJdG value. In this case the optimal configuration is obtained at $m_L = 0$, i.e. not a single vortex penetrates the cylinder. Note also that in Fig. \ref{fig-dfdx} the $m_L=0$ state also gives a field we associate with $h_{c_3}$ but a lower one than for $m_{L} = 1$ which agrees with the vortex expulsion from the nano-island at lower fields. It should be borne in mind that this agreement is only qualitative as we have solved the linearized GL equation that is in principle not justified below $H_{c3}$ where robust superconductivity develops.

\section{Transport effects near $H_{c3}$}\label{Sec:Transport}

Onset of superconducting transition, either near the critical temperature $(T_c)$ or critical field $(H_{c2})$, gives rise to strong corrections to normal-state properties, e.g., conductivity, due to Aslamazov-Larkin, Maki-Thompson, and density of states fluctuation effects [\onlinecite{Varlamov}]. In the context of surface superconductivity corresponding effects were considered for transport coefficients [\onlinecite{Schmidt,Zyuzin}] and thermodynamic quantities [\onlinecite{Aleiner}] near $H_{c3}$. In particular, it was shown in Ref. [\onlinecite{Aleiner}] that surface states lead to peculiar magneto-oscillations even in a singly connected conductors. This interference mechanism of magneto-oscillations can be understood semiclassically from the bouncing orbits confined to the surface. Because superconducting fluctuation corrections are singular they exceed usual mesoscopic interference effects near $T_c$, making them accessible in experiments. The most widely studied geometry involves superconducting rings and cylinders [\onlinecite{Zadorozhny,Liu,Wang,Staley}]. The theory of nonlocal response near $T_c$ in a superconducting ring was put forward a long time ago [\onlinecite{Glazman}]. It is natural to expect, that given the localization property of surface superconductivity to the edge of the sample, the results of Ref. [\onlinecite{Glazman}] will be applicable to cylinders in fields near $H_{c3}$. This motivates us to investigate the manifestation of surface effects on magnetoconductance. We indeed find that an analogy exists to a ring geometry, and surface superconductivity gives rise to oscillatory magneto-conductance, however, there are subtle differences that deserve detailed explanation.              

\subsection{Qualitative discussion of periodicity} 

Lets us first determine the periodicity in the field. This can be seen from the condition \eqref{GL-hc3-cond} with $h_{c3}$ replaced by some field above it, $h > h_{c3}$,
\begin{equation}\label{AB-cond}
\sqrt{ 2 m_L / h }  = x_R - x_0
\end{equation}
This time, we look at this equation as the condition for optimal combination of $m_L$ and $h$. Now imagine that we are at local maximum (of some quantity, e.g., conductivity)  so that for some value of $m_L = m_{L,0}$ there is a field $h_0$ such that the relation \eqref{AB-cond} is satisfied, $\sqrt{ 2 m_{L,0} / h_{0} } = x_R - x_0$. Now lets change $m_{L,0} \rightarrow m_{L,0} + 1$ and see what is the change $\delta h$ of $h$ such that the above condition still holds, namely, that $\sqrt{ 2 (m_{L,0}+1) / (h_{0} + \delta h)}  = x_R - x_0$. These two last conditions can be equivalently written as $m_{L,0} = \frac{1}{2} h_0  (x_R - x_0)^2$, and $m_{L,0} + 1 = \frac{1}{2}(h_0 + \delta h)  (x_R - x_0)^2$, which means that $\delta h = \frac{ 2 }{ (x_R - x_0)^2 }$ or in original units, $\delta H = \frac{ 2 H_{c2} \xi^2_{GL} }{(R - \delta R)^2}$, so that 
\begin{equation}\label{AB-cond_1}
\delta H \pi (R - \delta R)^2 = 2 \pi H_{c2} \xi^2_{GL} = \Phi_0\, .
\end{equation}
Thus, the periodicity is determined by the total flux enclosed by the surface states being a multiple of flux quantum $\Phi_0$. 
When the ratio of the flux enclosed by the surface state $\Phi_R'$ to $\Phi_0$, $\Phi_R' / \Phi_0 = \frac{1}{2} h (x_R - x_0)^2$ is integer the (integer) optimal angular momentum giving the fluctuation with the lowest energy is $m_L  = \Phi_R' / \Phi_0$.
For generic magnetic field the optimal angular momentum is
\begin{equation}\label{m-opt}
\bar{m}_L = \mathbb{I}[\Phi_R' / \Phi_0]\, ,
\end{equation}
where $\mathbb{I}[y]$ is the best integer approximation to $y$.  
Equation \eqref{m-opt} is another expression of flux periodicity.
We show in the next subsections that the periodicity of the fluctuation spectrum gives rise to Aharonov-Bohm oscillations of conductance promoted by surface superconductivity fluctuations. 

\subsection{Time-dependent GL theory}

To describe time-dependent fluctuations of superconductivity at the onset of transition we employ time-dependent Ginzburg-Landau (TDGL) equation. In the dimensionless notations of Eq. \eqref{GL-A} it takes the form  
\begin{equation}\label{TDGL-1}
-\frac{1 }{ x} \partial_x \left( x \partial_x f \right) + \frac{m_L^2}{ x^2 } f  - h m_L f + \frac{h^2}{4} x^2 f - f = - \tau_{GL}  \partial_t f,
\end{equation}
where $\tau_{GL}=\gamma/|\alpha|$ is the GL relaxation time. This equation has to be solved with the same boundary condition as before. Generically one solves for the spectrum of excitations, $\phi_{n,m_L}(x) e^{ i m_L \phi }$, satisfying 
\begin{equation}\label{TDGL-2}
\left[- \frac{1 }{ x} \partial_x  x \partial_x + \frac{m_L^2}{ x^2 }  - h m_L + \frac{h^2}{4} x^2  - 1\right]\phi_{n,m_L} = \lambda_{n,m_L} \phi_{n,m_L},
\end{equation}
so that the relaxation dynamics of fluctuations is given in terms of the expansion,
\begin{equation}\label{Psi-c-phi}
\Psi (x,t) = \sum_{n,m_L} c_{n,m_L} (t) \phi_{n,m_L}(x)
\end{equation}
with the wave functions $\phi_{n,m_L}(x)$ normalized to one, and expansion coefficients decaying exponentially 
\begin{equation}
c_{n,m_L}(t) = c_{n,m_L}(0)\exp(-t \lambda_{n,m_L}/\tau_{GL}).  
\end{equation}
In contrast to the problem of fluctuation in a ring near $T_c$ considered in Ref. [\onlinecite{Glazman}], in our problem $\tau_{GL}$ is regular upon approaching the $h_{c3}$ from above but the eigenvalues approach zero making the modes long lived instead. Our main conclusion here would be that at sufficiently low temperatures relatively large cylinders, $x_R \gg 1 $, behave essentially similar to the ring.
As in the cylinder, the condensate wave function is not confined in the radial direction; this requires a separate analysis. To see how the above analogy comes about, we will need more detailed information on the excitation spectrum $\lambda_{n,m_L}$. It is comprised of bands labeled by $n$ and index $m_L$ in quasi-continuum. 

For large cylinders and large angular momenta, the same reasoning that allowed us to replace Eq.~\eqref{GL-A} with Eq.~\eqref{GL-1D} also leads us to the approximate Eq.~\eqref{TDGL-2}:
\begin{equation}\label{TDGL-1D}
\left[-  \partial_x^2 + h^2 ( x - x_{\min})^2 - 1  \right] \phi_{n,m_L}  = \lambda_{n,m_L} \phi_{n,m_L}\, ,
\end{equation}
where the potential minimum is defined by Eq.~\eqref{x-min}.
Again, Eq. \eqref{TDGL-1D} is solved with the boundary conditions $[\partial_{x} \phi_{n,m_L}(x)]_{x =x_R} = 0$. In this approximation, 
\begin{equation}
\lambda_{n,m_L} = \lambda_{0,m_L} + 2 h n,
\end{equation}
and we focus for now on the lowest subband, $n =0$ in the limit of $x_R \gg 1$. 
We will aim at approximate expression for $\lambda_{0,m_L}$ that would suffice for our purposes.  
The approximate spectrum of excitations has to satisfy the following properties: (1) $\lambda_{0,m_L} = h- 1$ for $x_{\min} = x_R$; (2) $\lambda_{0,m_L} = h - h_{c3}$ for $x_{\min} = x_R - x_0$, where $x_0 \approx 0.59$ as in the semi-infinite geometry; (3) $\lambda_{0,m_L} \approx h - 1$ for $x_{\min} \lesssim x_R - 2 x_0$, since far from the boundary the boundary condition plays no role; (4) $\lambda_{0,m_L} \gg h - 1$ for $x_{\min} \gtrsim x_R + x_0$. To satisfy these conditions at the minimal level we adopt the following approximation:
\begin{equation}\label{lambda-xmin}
\lambda_{0,m_L}(x_{\min}) \approx h - h_{c3} + \frac{ h_{c3} - 1}{ x_0^2 } [x_{\min}- (x_R - x_{0}) ]^2,
\end{equation}
which adequately describes the excitations localized within a few $\xi_{GL}$ distance from the cylinder's edge.
The main feature of the approximate equation \eqref{lambda-xmin} is that $\lambda_{0,m_L}(x_{\min})$ has a minimum at $x_{\min} = x_R - x_0$ and Eq. \eqref{lambda-xmin} can be viewed as the expansion of $\lambda_{m_L,0}(x_{\min})$ around the minimum. The surface superconductivity exists because of this minimum and the importance of the latter is not unexpected.

The approximation \eqref{lambda-xmin} is only meaningful if the excitations contributing to the observable in question reside close to the ring of radius $x_R - x_0$. We will check this in each case. The energies are functions of the $m_L$ quantum number rather than on auxiliary parameter, $x_{\min}$. 
Substituting Eq. \eqref{x-min} into \eqref{lambda-xmin} we get explicitly, 
\begin{equation}\label{lambda}
\lambda_{0,m_L}(x_{\min}) \approx h - h_{c3} + \frac{ h_{c3} - 1}{ x_0^2 } \left[\sqrt{ 2 m_L / h}- (x_R - x_{0}) \right]^2 \, .
\end{equation}
We give a sufficient condition on temperature and field for this to be justified.
It follows from the constraint on the thermal occupation of different excitations.

The occupation number of an excitation of energy $\epsilon_{n,m_L}$ is determined by the Bose function 
$N( \epsilon ) = \left[ \exp\left(\epsilon /T \right)- 1 \right]^{-1}$.
As Eq.~\eqref{GL-A} is Eq.~\eqref{GL-Eq} divided by $|\alpha|$ the energies of excitations in the original units are 
$\epsilon_{n,m_L} = |\alpha| \lambda_{n,m_L} $.
Therefore introducing the thermal de Broglie wavelength, $l_{T} = \sqrt{1/ ( 2 m T)}$, we obtain
$N( \lambda_{0,m_L} ) = \left[ \exp\left( l_T^2 \lambda_{0,m_L}/ \xi^2_{GL}  \right)- 1 \right]^{-1}$.
In order to have many fluctuations around the ring $x_R - x_0$ without exciting too many fluctuations outside of it we impose a condition
\begin{equation}
    \frac{l_T^2 }{ \xi^2_{GL} }  (h - h_{c3})  \ll 1, \quad  \frac{l_T^2 }{ \xi^2_{GL} }  (h - 1)  \gg 1.
\end{equation}
These conditions require (returning to the original units)
\begin{equation}
H_{c3} - H_{c2}  \gg H_{c2} \frac{ \xi^2_{GL} }{l_T^2 }\, ,
\end{equation}
which in turn means that the temperature should be reasonably low such that $\xi^2_{GL}/l_T^2 \lesssim 1$. 

We now bring our model of the fluctuation spectrum [Eq. \eqref{lambda}] to the form that is most suitable for the computation of the current and conductivity. To this end, we make an expansion of the model energy [Eq.~\eqref{lambda-xmin}] around the optimal angular momentum $\bar{m}_L$ defined by Eq.~\eqref{m-opt} relying on the smallness of $1/\bar{m}_L$ in the limit of large cylinder,
\begin{equation}\label{exp-opt1}
\sqrt{ 2 m_L/h} - (x_R - x_0) = \sqrt{ 2 \bar{m}_L/h} +  \frac{\sqrt{ 2 /h}}{2\sqrt{\bar{m}_L}} \Delta m  - (x_R - x_0)\, ,
\end{equation}
where $\Delta m = m_L - \bar{m}_L \ll \bar{m}_L$ may still be large.
As the fractional part of the flux ratio,  $\mathbb{F}[\Phi'_R/\Phi_0] = \Phi'_R/\Phi_0 - \mathbb{I}[\Phi'_R/\Phi_0]$ is much smaller than its integer part  $\mathbb{I}[\Phi'_R/\Phi_0] $ we write
\begin{align}\label{exp-opt2}
&\sqrt{ 2 \bar{m}_L/h} - (x_R - x_0) = \sqrt{ (2 /h) (\Phi'_R/ \Phi_0 - \mathbb{F}[\Phi'_R/ \Phi_0] )} \nonumber 
\\ & - (x_R - x_0) \approx -\sqrt{ (2 /h)} \frac{\mathbb{F}[\Phi_R/ \Phi_0]}{2\sqrt{\Phi'_R/ \Phi_0}},
\end{align}
so that the expression \eqref{exp-opt1} becomes,
\begin{equation}\label{exp-opt3}
\sqrt{ 2 m_L / h}- (x_R - x_{0}) \approx -\sqrt{ (2 /h)} \frac{\mathbb{F}[\Phi_R/ \Phi_0]}{2\sqrt{\Phi'_R/ \Phi_0}}+  \frac{\sqrt{ 2 /h}}{2\sqrt{\bar{m}_L}} \Delta m. 
\end{equation}
As the magnetic flux contains the large number of flux quanta, we can set in \eqref{exp-opt3} based on the definition \eqref{m-opt},
$\bar{m}_L = \mathbb{I}[\Phi_R' / \Phi_0] \approx \Phi_R' / \Phi_0$, and obtain for the energies, 
\begin{align}\label{exp-opt4}
\lambda_{0,m_L}  \approx h - h_{c3} + \frac{ h_{c3} - 1}{2 h x_0^2 } \frac{\Phi_0}{\Phi'_R} [\mathbb{F}[\Phi_R/ \Phi_0] - \Delta m ]^2. 
\end{align}
The term $[\mathbb{F}[\Phi'_R/ \Phi_0] - \Delta m ]^2$ reflects the periodicity of the spectra of excitations with flux. As the flux increases the optimal angular momentum $\bar{m}_L$ also increases in step like fashion. 
This expression vanishes in the middle of each such step.
This periodic behavior is similar to that in the ring.
The periodicity of the spectrum overall is not perfectly periodic as the amplitude of these oscillations,  $\frac{ h_{c3} - 1}{2 h x_0^2 }$ is itself a slow function of the field.
More importantly, the spectrum \eqref{exp-opt4} softens towards the transition at $h = h_{c_3}$.
Neglecting the field dependence of the amplitude, and setting $\Phi'_R \approx \Phi_R$ in the limit of large cylinder, we have for the reduced energy,
\begin{equation}\label{l-bar}
\bar{\lambda}_{m_L}=\frac{\lambda_{0,m_L}}{h-h_{c3}}, \, \,\,
\bar{\lambda}_{m_L} = 1 +\frac{ C(\xi_{GL}/R)^2 }{(h - h_{c_3} )} [\mathbb{F}[\Phi_R/ \Phi_0] - \Delta m ]^2, 
\end{equation}
where $C$ is a numerical constant of the order one. The analogy to the one-dimensional (1D) geometry of the ring considered in Ref. [\onlinecite{Glazman}] becomes complete if we define
\begin{equation}\label{xi-H}
\xi_{H}^2 = C \frac{\xi_{GL}^2}{h - h_{c3}}
\end{equation}
so that the reduced energies become
\begin{equation}
\bar{\lambda}_{m_L} = 1 +  \frac{ \xi_{H}^2}{ R^2} [\mathbb{F}[\Phi_R/ \Phi_0] - \Delta m ]^2.
\end{equation}
Following the analysis of the ring carried in Ref. [\onlinecite{Glazman}] we conclude that the diamagnetic response of the cylinder is dominated by the superconducting fluctuations at its circular edge boundary for the range of fields $H-H_{c_3} \ll  H_{c_3} - H_{c_2}$, and not for too low temperatures. The main conclusion we draw from this analysis is that the fluctuation part of the response is divergent and oscillates with the Aharonov-Bohm periodicity determined by the cross-sectional area of the cylinder. As long as the radius of the cylinder is larger than the zero-field coherence length, the oscillations are predicted along with the critical enhancement of the diamagnetic response. This is explicitly demonstrated in the next section where we compute correction to the conductivity from the superconducting fluctuations at the edge. 

\subsection{Magnetoconductivity oscillations}

We calculate the fluctuation correction to the conductivity with the help of the Kubo formula
\begin{equation}\label{d-sigma}
\delta\sigma_{\alpha\beta}=  \frac{ 2}{ T} \lim_{\omega \rightarrow 0} \int_{-\infty}^{\infty} d t \cos(\omega t) \langle J_{\alpha}(\bm{r},t) J_{\beta}(\bm{r}',0) \rangle 
\end{equation}
which relates the elements of the conductivity tensor to the correlation function of the components of the operator for the supercurrent from Eq. \eqref{GL-J}. 
We will be interested in the fluctuation conductivity along the edge, and will focus on the azimuthal components of the current operator,
$J(\bm{r}) = \hat{\varphi} \cdot \bm{J}(\bm{r})$.
Next we rewrite the azimuthal component of the supercurrent, Eq. \eqref{GL-J}, in terms of mode expansion, Eq. \eqref{Psi-c-phi}. In the dimensionless coordinates it takes the form
\begin{align}\label{J-az}
J(x,\varphi,t)  = \frac{e}{m\xi_{GL}}  \sum_{n,m_L}  \sum_{n',m'_L} 
c^*_{n,m_L}(t) c_{n',m'_L}(t)\nonumber \\ 
\left[\phi^*_{n',m_L'}(x,\varphi)\left(\frac{m_L + m'_L}{x} - h x  \right) \phi_{n,m_L}(x,\varphi)  \right]\, .
\end{align}
To simplify the discussion we will focus on the contribution under the conditions specified in the previous section only the lowest band, is appreciably occupied.
We therefore include only the lowest mode $n=n'=0$ contribution in Eq.~\eqref{J-az}. Also to simplify notations, we will suppress index zero corresponding to $n=0$ from all the quantities and use a shorthand notation for the angular momentum,
$m_L \to L$ so that in these notations,  $c_{0,m_L}\to c_{L}$, $\phi_{0,m_L}\to \phi_{L}$, etc. 
The current correlation function reads as
\begin{align}\label{JJ}
\langle J(t) J(0) \rangle & = 
\left(\frac{ e}{m \xi_{GL}} \right)^2 \sum_{L,L',M,M'} 
\langle c^*_{L}(t) c_{L+M}(t) c^*_{L'}(0) c_{L'+M'}(0) \rangle
\notag \\
\times & 
\left[\phi^*_{L}(x,\varphi)\left[\frac{2L+M}{x} - h x  \right] \phi_{L+M}(x,\varphi)  \right]
\notag \\
\times & 
\left[\phi^*_{L'}(x',\varphi')\left[\frac{2L' + M'}{x'} - h x'  \right] \phi_{L'+M'}(x',\varphi')  \right].
\end{align}

We now introduce the total current flowing across the radial cross-section $I = \int_{0}^{R} d \rho J(\rho)$. 
As the wave functions are normalized to $ \int_0^R d \rho  \rho |\psi(\rho)|^2 = 1$ which then gives the normalization condition,
$\int_0^{x_R} d x x |\phi_{L,n}(x)|^2 = \xi_{GL}^{-2}$.

Since only  the states localized at the edge of the cylinder make an appreciable contribution to the conductivity the matrix elements in Eq. \eqref{JJ} only weekly depend on the quantum number $M$, 
we can write
\begin{align}\label{ME-J}
&\int_0^R d r \phi^*_{L}(x,\varphi)\left[\frac{2L+M}{x} - h x  \right] \phi_{L+M}(x,\varphi) \nonumber \\ 
&\approx
e^{ i M \varphi} \xi_{GL} \int_0^{x_R} d x |\phi_{L}|^2 \left[\frac{2L+M}{x'_R} - h x'_R  \right] 
\end{align}
where $x'_R$ differs from $x_R$ by a number of order of unity. 
To fix $x'_R$ we notice that the magnetic flux is thermodynamically conjugated to the current, and at the local minima of the energy as a function of flux the current vanishes.
The consistency with the model dispersion hence is achieved if we identify $x'_R$ with $x_R - x_0$ introduced earlier to parametrize the spectrum of excitations [Eq.~\eqref{lambda}]. 
Indeed in this case the diagonal matrix element of the current, Eq.~\eqref{ME-J} vanishes in the states with the angular momentum $L$ producing the minimum in energy.
More precisely, if for some state characterized by an integer $L$ the energy given by Eq.~\eqref{lambda} vanishes, and the current expectation value in such state vanishes as well.   
Arguing as before we obtain the following estimate: 
\begin{align}
&\int_0^R d r \phi^*_{L}(x,\varphi)\left[\frac{2L+M}{x} - h x \right] \phi_{L+M}(x,\varphi)\nonumber \\  
&\approx
\frac{\xi_{GL}}{R^2} ( 2 \Delta L + M - 2 \mathbb{F}[\Phi_R/\Phi_0]). 
\end{align}
Substituting these estimate of the current matrix elements back into Eq. \eqref{JJ}, and using the definition of $I$ we obtain, for a measure of the fluctuation superconductivity,
\begin{align}\label{II}
\langle I(t) I(0) \rangle = 
\left(\frac{e}{m R^2} \right)^2 & \sum_{\Delta L,M} e^{iM(\varphi-\varphi')}
\langle c^*_{L}(t) c_{L+M}(t) c^*_{L+M}(0) c_{M}(0) \rangle \nonumber \\ 
& \times(2\Delta L + M - 2\mathbb{F}[\Phi_R/\Phi_0])^2\, ,
\end{align}
where again $\Delta L = L - \bar{L}$ is the deviation of the angular momentum from its optimal value,
$\bar{L} = \mathbb{I}[\Phi'_R/\Phi_0] \approx \mathbb{I}[\Phi_R/\Phi_0]$.
It is henceforth possible to rewrite the combination $2\Delta L + M - 2\mathbb{F}[\Phi_R/\Phi_0]$ simply as
$2L + M - 2\Phi_R/\Phi_0$.

Next, we discuss the thermal averages appearing in Eq.~\eqref{II}.
Assuming, as usual in the TDGL approach the different fluctuation modes are Gaussian and statistically independent, we obtain
\begin{align}
\langle c^*_{L}(t) c_{L'}(0) \rangle  = \delta_{L,L'}  \frac{ T}{|\alpha| \lambda_{L} }\exp(-t \lambda_{L}/\tau_{GL})\, .
\end{align}
It is important to stress that we do not specify the time $\tau_{GL}$ and in contrast to the standard case of 
zero-field fluctuation corrections, this time scale is not divergent as $H$ approaches $H_{c_3}$.
The life time of fluctuations does become infinite upon the transition thanks to the softening of the excitation spectrum. It follows that, for $M \neq 0$,
\begin{align}\label{4-corr}
&\langle c^*_{L}(t) c_{L+M}(t) c^*_{L'}(0) c_{L'+M'}(0) \rangle =
\notag \\
& = \delta_{L,L'+M'} \delta_{M',-M} \langle c^*_{L}(t) c_{L}(0) \rangle \langle c_{L+M}(t) c^*_{L+M}(0) \rangle
\notag \\
& = \delta_{L,L'+M'} \delta_{M',-M} \frac{ T}{|\alpha| \lambda_{L} } \frac{ T}{|\alpha| \lambda_{L+M} }
\notag \\
& \times \exp 
\left[ - \frac{t}{\tau_{GL} }\left(\lambda_{L}  + \lambda_{L+M} \right)\right]\, .
\end{align}
The extra contribution to the correlation function, \eqref{4-corr} obtained for $M =0$ is time independent, and therefore excluded from the final result due to the limiting procedure implied by Eq.~\eqref{d-sigma}. 

\begin{figure}
\includegraphics[width=0.4\textwidth]{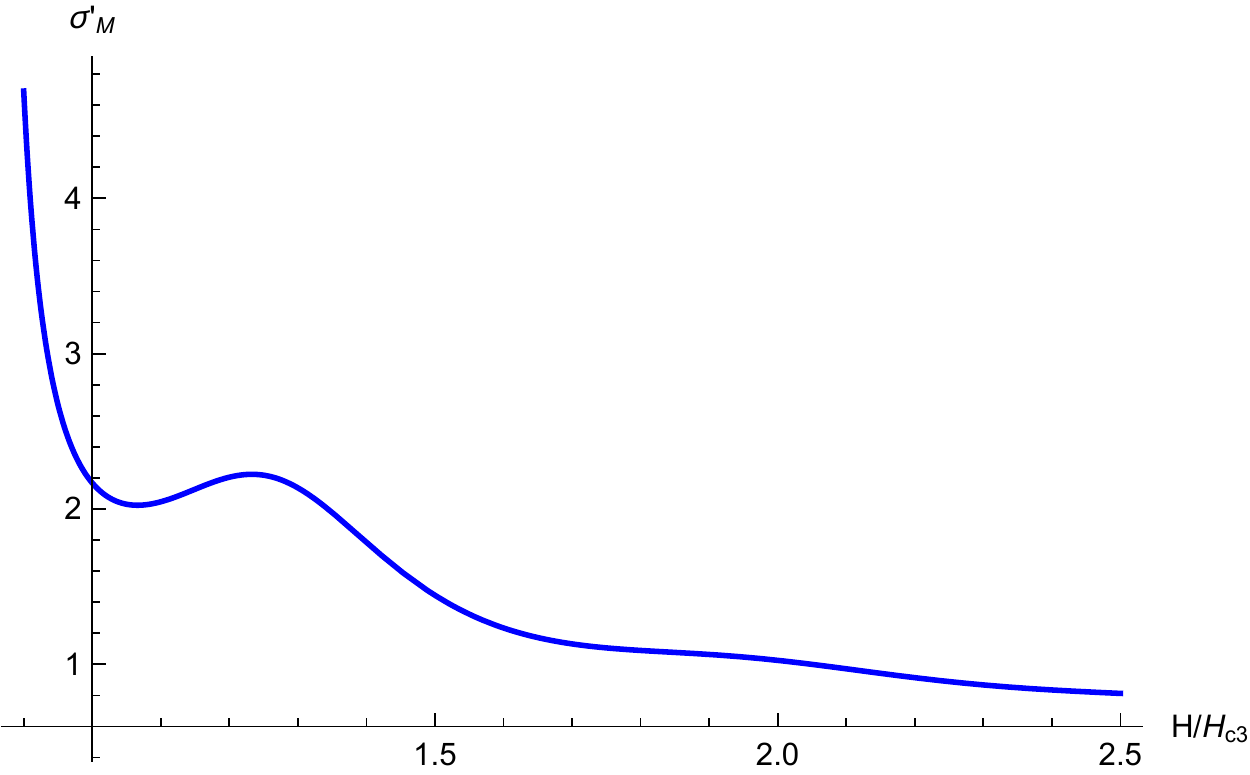}
\caption{
The Fourier component of the conductivity, $\sigma_M'$ as a function of the magnetic field (arbitrary units) plotted by using Eq.~\eqref{res}. As the critical field $H_{c3}$ is approached from above the conductivity diverges while exhibiting Aharonov-Bohm oscillations. The radius of the cylinder is taken to be $R/\xi_{GL} \approx 4.2$.} 
\label{Fig:Osc}
\end{figure}

Repeating the line of arguments in Ref. [\onlinecite{Glazman}], we introduce the Fourier components of the conductivity,
\begin{align}\label{FC}
\delta \sigma = \frac{1 }{ 2 \pi R} \sum_m e^{ i m (\varphi - \varphi')} \sigma'_{m}
\end{align}
to characterize the non-local transport. Substituting Eqs.~\eqref{II} and \eqref{4-corr} into Eq.~\eqref{d-sigma}, we obtain
\begin{align}\label{sigma-M}
\sigma'_M = 4 \pi R T \left(\frac{e}{m R^2} \right)^2 \frac{\tau_{GL}}{|\alpha|^2}\sum_L \frac{( 2 L + M - 2 \Phi_R/\Phi_0)^2 }{\lambda_L \lambda_{L+M} ( \lambda_{L} + \lambda_{L+M} )}
\end{align}
We further note that
\begin{align}
&  2 L + M - 2 \Phi_R/\Phi_0 
 \notag \\
& =  
 \frac{1}{M^2}
 \left[ (L + M -  \Phi_R/\Phi_0)^2 - (L  -  \Phi_R/\Phi_0)^2 \right]\, ,
\end{align}
which according to the definitions \eqref{l-bar} and \eqref{xi-H} allows  us to write, 
\begin{align}
(2 L + M - 2 \Phi_R/\Phi_0 )^2 =  \frac{1}{M^2} \left( \frac{ R }{ \xi_{H}} \right)^4 \left( \bar{\lambda}_{L +M} - \bar{\lambda}_{L} \right)^2.
\end{align}
Noticing the relation \eqref{l-bar}, 
$\bar{\lambda}_{L}=\lambda_{L}/(h-h_{c3})$ and applying the algebraic identity,
\begin{align}
\frac{(\bar{\lambda}_{L+M} - \bar{\lambda}_{L})^2}{(\bar{\lambda}_{L+M} + \bar{\lambda}_{L})\bar{\lambda}_{L}\bar{\lambda}_{L+M}}
=
\frac{1}{\bar{\lambda}_{L}} + \frac{1}{\bar{\lambda}_{L+M}} - \frac{ 4 }{\bar{\lambda}_{L} +\bar{\lambda}_{L+M}}
\end{align}
we obtain from Eq.~\eqref{sigma-M} 
\begin{align}\label{series}
\sigma'_M = \frac{ e^2 }{ 2 }\frac{T \tau_{GL}}{ h - h_{c_3}} \frac{ R}{ M^2 } 
\sum_L \left[\frac{ 1 }{ 1 + (\xi_H/R)^2 (L  - \Phi_R/\Phi_0)^2 } \right. \nonumber \\ -\left. 
\frac{2}{2 +  (\xi_H/R)^2 \{ (L  -\Phi_R/\Phi_0)^2  + (L + M  - \Phi_R/\Phi_0)^2 \} }
\right]\, .
\end{align}
The presence of the temperature in the final result reminds us that we discuss the effect of classical fluctuations. Although the last expression differs from the one obtained for the ring geometry in Ref. [\onlinecite{Glazman}] [see their Eq. (18)] as it implicitly assumes the limitation on the possible angular momentum, $\Delta L = L - \bar{L} \ll \bar{L}$, it leads nevertheless to the same expression for the Fourier components of the conductivity for the large range of Fourier components satisfying $M \lesssim \bar{L}  \approx \Phi_R/\Phi_0$ as the series in \eqref{series} rapidly converges [Fourier components decay as $1/L^2$]. For the same reason, the summation in \eqref{series} can be extended to all angular momenta.
Summing the series and retaining only the most divergent term in $h-h_{c3}$, we find for the conductivity our main result
 \begin{equation}\label{res}
\delta \sigma \simeq
e^2\left(\frac{T \tau_{GL}}{ h - h_{c_3}}\right)  \left(\frac{ R^2}{ \xi_H}\right)
\frac{\sinh(2 \pi R / \xi_H)}{\cosh(2 \pi R / \xi_{H}) - \cos (2 \pi \Phi_R/\Phi_0) }\, .
\end{equation}
In the immediate vicinity of the critical field, $R/\xi_{H}\ll1$, conductivity correction diverges as $\delta\sigma\propto 1/(H-H_{c3})$. 
This correction is less singular than the standard Aslamazov-Larkin correction in strictly one-dimensional open systems scaling as $\sim 1/(T- T_c)^{3/2}$ at zero magnetic field (see also Ref. [\onlinecite{Zyuzin}]). The difference originates from the geometry of our problem. As the radius of the cylinder becomes smaller than the correlation length, $\xi_{H}$ the conductivity $\sigma_M'$ corresponds to the local conductivity in an open system. In other words, to get our result from the standard one in one-dimensional system one has to perform the  Fourier transformation of the latter that is the integration over wave vectors.
Since the typical wave vectors are $\sim \xi_H^{-1}$ our final result scales as $\xi_H^{3}  \xi_H^{-1} \propto (H- H_{c3})^{-1}$. We show a representative plot of the fluctuation edge conductivity correction as given by \eqref{res} in Fig.~\ref{Fig:Osc}.

Oscillations survive even far away from the transition where $R/\xi_{c3}\gg1$, however oscillatory term is exponentially suppressed in that parameter range as $\sim e^{-2\pi R/\xi_{H}}\cos(2\pi\Phi_R/\Phi_0)$.   

\section{Summary and discussions}\label{Sec:SumDis}

In this paper we reported a number of results concerning the critical field of superconductivity nucleation close to the edge of the samples and surface superconductivity overall. In Sec. \ref{Sec:Eilenberger-Hc3} we applied semiclassical Eilenberger formalism to study critical field $H_{c3}$ as a function of temperature and sample purity. This enabled us to extend the seminal analysis of Saint James and de Gennes applicable near $T_c$ to the whole range of parameters relevant for numerous experiments. Our main finding in this part of the paper is that the $H_{c3}/H_{c2}$ ratio is a nonmonotonic function of temperature that has a maximum $\sim 2$ for a set of parameters corresponding to crossover between ballistic and diffusive scattering in the bulk of the superconductor. It is of interest to contrast these findings to results reported in Ref. [\onlinecite{Gorokhov}] where nonmonotonic behavior of $H_{c3}/H_{c2}$ ratio was found in the case of two-band superconductor in the diffusive limit based on Usadel equations. In this case the ratio is always below the SJdG limit so that multi-band effects seem to suppress the difference between critical fields in the strongly disordered case at intermediate temperatures. This calls for more studies of intertwined effects of disorder and multiple bands in applications to novel unconventional superconductors. 

In Sec. \ref{Sec:Boundary-SSC} we analyzed robustness of surface superconductivity against diffusive scattering off the boundary and found that an estimate of $H_{c3}$ remains above the SJdG limit. 

In Sec. \ref{Sec:Hc3-Disc} we applied GL formalism to study $H_{c3}$ and surface effects in the case of devices of circular geometry and applied our results to STS measurements [\onlinecite{Roditchev}] of single vortex nucleation in mesoscopic samples. 

Magnetoresistance experiments seem to suggest that surface superconductivity is the primary cause of broadening of superconducting transition in high magnetic field, as data $R(T,H)$ are often benchmarked by $H_{c3}(T)$, which motivated us to look at transport phenomena in Sec. \ref{Sec:Transport}. We found that at the onset of superconducting transition there exists a correction to conductance that oscillates with Aharonov-Bohm flux. Oscillatory flux dependence of the conductance in quantum wires was reported in multiple measurements  [\onlinecite{Dynes,Kwok,Shahar,Markovic}]. In particular, magnetotransport observations in Ref. [\onlinecite{Markovic}] were interpreted in terms of the Weber blockade theory due to vortex formation inside the sample proposed in Ref. [\onlinecite{Pekker}]. This theory was also used to interpret the observations of an earlier experiment [\onlinecite{Shahar}]. In contrast, our results support the point of view that oscillatory response of magnetoresistance is a much more general phenomenon which survives even in the limit when superconductivity is fully extinguished by high magnetic field.  

\section*{Acknowledgments} 

The authors are grateful to N. Katz, R. Prozorov, and especially M. Tsindlekht for fruitful discussions. We thank A. V. Andreev for communications regarding the results of Ref. [\onlinecite{Aleiner}].  

This work was financially supported by NSF Grant No. DMR-1606517 and in part by BSF Grant No. 2014107 (H. X.), by NSF CAREER Grant No. DMR-1653661 and Vilas Life Cycle Professorship program (A. L.), by BSF Grant No. 2016317 and by the Israel Science Foundation, Grant No. 1287/15 (M. K.). V. K. was supported by the U.S. Department of Energy, Office of Science, Basic Energy Sciences, Materials Sciences and Engineering Division. The Ames Laboratory is operated for the U.S. DOE by Iowa State University under Contract No. DE-AC02-07CH11358. M.K. and A.L. are grateful for the hospitality of the Kavli Institute for Theoretical Physics during the program on ``Intertwined Order and Fluctuations in Quantum Materials", where parts of this work were completed and supported in part by the NSF under Grant No. NSF PHY11-25915.

\appendix

\section{Integrals}  \label{appendix-integrals}

Within this section we provide integrals that lead to Eq. \eqref{IntEq-Gorkov} in the main text starting from Eq. \eqref{Hc2-Gorkov} and trial wave function of the form Eq. \eqref{Delta-Trial1}. After multiplication by $\Delta(x)$ the integral on the left-hand-side of Eq. \eqref{IntEq-Gorkov} is elementary as it is just a Gaussian: $L\int^{+\infty}_{-\infty}\Delta^2(x)dx=L\sqrt{\pi/2a}$. To calculate the rest integrals we introduce the hyperbolic coordinates
$x=r\cosh{\rho}, x'=r\sinh{\rho}$ with the Jacobian of transformation $dxdx'=rdrd\rho$. The off-diagonal term on the right-hand-side of Eq. \eqref{IntEq-Gorkov} reads
\begin{widetext}
\begin{align} 
&\iint_{-\infty}^{+\infty} \!\! K_1(x,x') \Delta(x) \left[ \Delta(x)-\Delta(x') \right]dx dx' = \frac{1}{2} \iint_{-\infty}^{+\infty} \!\! K_1(x,x')  \left[ \Delta(x)-\Delta(x') \right]^2 dx dx' = \frac{1}{4}\iint_{-\infty}^{+\infty}\! \frac{e^{-|x^2 -{x^\prime}^2|}}{|x -{x^\prime}|} \left( e^{-a x^2} - e^{-a {x^\prime}^2} \right)^2 dxdx' \nonumber \\
& = \frac{1}{2} \int_{0}^{\infty} \! d x \int_{-x}^{x} d x^\prime \left( \frac{1}{x+x^\prime} + \frac{1}{x-x^\prime} \right) e^{ {x'}^2 -x^2 }\left( e^{-a x^2} - e^{-a {x^\prime}^2} \right)^2 \nonumber \\ &= \int_{-\infty}^{+\infty} \! d\rho \cosh{\rho} \int_{0}^{\infty}\! dr \left\{ e^{-(1+2 a \cosh^2{\rho})r^2}  + e^{-(1+2 a \sinh^2{\rho})r^2} - 2 e^{-[(1+a)+2a\sinh^2{\rho}]r^2} \right\} \nonumber \\ 
&= \frac{\sqrt{\pi}}{2} \int_{-\infty}^{+\infty} \! d\rho \cosh{\rho} \left( \frac{1}{\sqrt{1+2 a \cosh^2\rho}} + \frac{1}{\sqrt{1+2 a \sinh^2\rho}} - \frac{2}{\sqrt{(1+a)+2 a \sinh^2{\rho}}}  \right) =  \frac{1}{2} \sqrt{\frac{\pi}{2 a}} \ln{\frac{(1+a)^2}{1+2 a}} .
\end{align}
In a similar fashion, the diagonal term reads
\begin{align}
\int_{-\infty}^{+\infty}\!\!\!K_0(x) \Delta^2(x) dx = 2\!\! \int_{0}^{\infty} \!\!\! dx\!\! \int_{0}^{x}\!\! d x' e^{-(2 a+1)x^2 + {x'}^2} x' \ln{ \frac{x+x^\prime}{x-x^\prime}} dx' = 4 \int_{0}^{\infty}\! d \rho \, \rho \, \sinh{\rho} \int_{0}^{\infty}\!dr  \, r^2 e^{-[(2a+1)\cosh^2{\rho} -\sinh^2{\rho}] r^2} \nonumber \\ =  \int_{0}^{\infty}\! \frac{\sqrt{\pi} \rho \sinh{\rho}}{\left(2a \cosh^2{\rho} +1\right)^{3/2}} d \rho = \frac{1}{2} \sqrt{\frac{\pi}{2a}}\ln  \frac{1 + 2a}{2 a}.
\end{align}
\end{widetext} 
Combining all three integrals together we arrive at Eq. \eqref{Hc2-Gorkov}. 

\section{Numerical algorithm} \label{appendix-algorithm}

For the numerical calculation of the lowest eigenvalue and corresponding eigenstate of the integral equation \eqref{IntEq} we change variable and cast the integral volume to finite interval  $x = f(\zeta)$ , $dx =  \mu(\zeta) d \zeta$, where we choose $f(\zeta) =  \frac{\zeta}{1-\zeta}$ and $\mu(\zeta) = \frac{1}{(1-\zeta)^2}$ with $\zeta \in [0,1]$. Then we discretize the integral according to the Newton-Cotes quadrature rules by choosing the representative points $\{ \zeta_i \}_{i=0}^{N}$, $\zeta_i = is+\zeta_0$, $s = (1-\zeta_0)/N$, where $\zeta_0=0$. When $N \gg 1$, Eq.~\eqref{IntEq} is approximated by the matrix equation
\begin{equation} 
L^{(N)} D_i^{(N)} = \sum_{j=0}^{N} H_{ij}^{(N)} D_j^{(N)},
\end{equation}
with $D_i^{(N)} = \sqrt{\mu_i} \Delta(x_i)$ and the discretized kernel 
\begin{eqnarray} 
&& H_{ij}^{(N)} =  
 \delta_{ij} \Big[ K_0(x_i) + \sum_{\substack{j \neq i}}^{N} \mu_j K_1(x_i,x_j) \Big]\nonumber \\ 
 && -(1-\delta_{ij}) \sqrt{\mu_i \mu_j } K_1(x_i,x_j)
- \sqrt{\mu_i \mu_j} K_{2}(x_i, x_j), 
\end{eqnarray}
where $x_i = f(\zeta_i)$ and $\mu_i = w_i \mu(\zeta_i)$ with $w_i = s/2$ for $i \in \{0, N\}$ and $w_i = s$ for $0 < i <N$ being the Newton-Cotes weights. The discretized kernel for the diffusive case was used to produce results reported in Fig. \ref{fig-Hc3-Diffusive}(b).


\end{document}